\documentclass[journal]{IEEEtran}

\usepackage[numbers,sort&compress]{natbib}
\usepackage{graphicx}
\usepackage{multirow}
\usepackage{amsmath,amssymb,amsfonts}
\usepackage{amsthm}
\usepackage{mathrsfs}
\usepackage{xcolor}
\usepackage{textcomp}
\usepackage{booktabs}
\usepackage{algorithm}
\usepackage{algorithmicx}
\usepackage{algpseudocode}
\usepackage{listings}
\usepackage{makecell}
\usepackage{array}
\usepackage{tabularx}
\usepackage{hyperref}
\newcolumntype{L}[1]{>{\raggedright\arraybackslash}p{#1}}

\theoremstyle{plain}

\theoremstyle{definition}

\theoremstyle{remark}

\begin{document}

\title{A Survey of Zero-Knowledge Proof Based Verifiable Machine Learning}

\author{Zhizhi Peng, Chonghe Zhao, Taotao Wang, Guofu Liao, Zibin Lin, Yifeng Liu, Bin Cao, Long Shi, Qing Yang, and Shengli Zhang%

\thanks{Zhizhi Peng, Taotao Wang, Guofu Liao, Zibin Lin, Qing Yang, and Shengli Zhang are with the College of Electronics and Information Engineering, Shenzhen University, Shenzhen, Guangdong Province, China. E-mail: p1878575@163.com; ttwang@szu.edu.cn; liaoguofu2022@email.szu.edu.cn; linaacc9595@gmail.com; yang.qing@szu.edu.cn; zsl@szu.edu.cn.}%
\thanks{Chonghe Zhao is with the School of Computer Science and Cyber Engineering, Guangzhou University, Guangzhou, Guangdong Province, China. E-mail: chonghe.zhao@gzhu.edu.cn.}%
\thanks{Yifeng Liu is with the School of Electronic and Information Engineering, South China University of Technology, Guangzhou, Guangdong Province, China. E-mail: 13539213368@163.com.}%
\thanks{Bin Cao is with the State Key Laboratory of Networking and Switching Technology, Beijing University of Posts and Telecommunications, Beijing, China. E-mail: caobin@bupt.edu.cn.}%
\thanks{Long Shi is with the School of Electronic and Optical Engineering, Nanjing University of Science and Technology, Nanjing, Jiangsu Province, China. E-mail: slong1007@gmail.com.}%
\thanks{\em{Corresponding author: Taotao Wang (ttwang@szu.edu.cn)}. }%
}

\maketitle

\begin{abstract}
Machine learning is increasingly deployed through outsourced and cloud-based pipelines, which improve accessibility but also raise concerns about computational integrity, data privacy, and model confidentiality. Zero-knowledge proofs (ZKPs) provide a compelling foundation for verifiable machine learning because they allow one party to certify that a training, testing, or inference result was produced by the claimed computation without revealing sensitive data or proprietary model parameters. Despite rapid progress in zero-knowledge machine learning (ZKML), the literature remains fragmented across different cryptographic settings, ML tasks, and system objectives. This survey presents a comprehensive review of ZKML research published from June 2017 to August 2025. We first introduce the basic ZKP formulations underlying ZKML and organize existing studies into three core tasks: verifiable training, verifiable testing, and verifiable inference. We then synthesize representative systems, compare their design choices, and analyze the main implementation bottlenecks, including limited circuit expressiveness, high proving cost, and deployment complexity. In addition, we summarize major techniques for improving generality and efficiency, review emerging commercial efforts, and discuss promising future directions. By consolidating the design space of ZKML, this survey aims to provide a structured reference for researchers and practitioners working on trustworthy and privacy-preserving machine learning.
\end{abstract}

\begin{IEEEkeywords}
Zero-knowledge Proof, Machine Learning, Verifiability, Model Security, Data Privacy
\end{IEEEkeywords}

\section{Introduction}\label{s:intro}
Artificial intelligence (AI), with machine learning (ML) as its core engine, is increasingly embedded in modern applications such as content generation, software engineering, finance, healthcare, and intelligent decision support. This progress is driven by rapid advances in model architectures, optimization methods, and training infrastructure. At the same time, the computational cost of state-of-the-art ML continues to rise sharply: larger models require larger datasets, more memory, and increasingly specialized hardware. Commercial systems such as ChatGPT~\cite{wu2023brief} and Midjourney~\cite{borji2022generated} exemplify this trend, as their performance depends not only on algorithmic innovations but also on large-scale investments in data and compute resources.

This resource asymmetry has accelerated the adoption of Machine Learning as a Service (MLaaS)~\cite{ribeiro2015mlaas}. In the MLaaS paradigm, clients outsource training, testing, or inference to external service providers that host the required models, hardware, or data-processing pipelines. While this model improves accessibility, it also creates two intertwined concerns. First, clients may be unwilling to expose sensitive data to external providers because of privacy risks and frequent real-world data breaches. Second, service providers often regard trained models as valuable intellectual property and are reluctant to reveal model parameters or internal execution details. At the same time, clients still need assurance that the returned model, metric, or prediction was produced by the declared computation rather than fabricated, shortcut, or generated using an unintended model or dataset. These tensions make trust and privacy central challenges in outsourced ML~\cite{al2019privacy,wang2019security}.

These requirements motivate verifiable machine learning, whose goal is to establish both computational integrity and privacy. Zero-knowledge proof (ZKP) is particularly well suited to this objective. A ZKP allows one party to convince another party that a statement is true without revealing any information beyond the validity of that statement~\cite{de1992zero}. In the ZKML setting, the statement can encode claims such as ``this prediction equals the output of the committed model on the declared input'' or ``these updated parameters were produced by the declared training procedure on the committed dataset.'' By coupling ML computation with cryptographic proof generation, zero-knowledge machine learning (ZKML) enables mutually distrustful parties to establish trust in an outsourced ML task without disclosing sensitive training data, inference inputs, or proprietary model parameters.

\begin{figure*}[t]
	\centering
	\includegraphics[width=\linewidth]{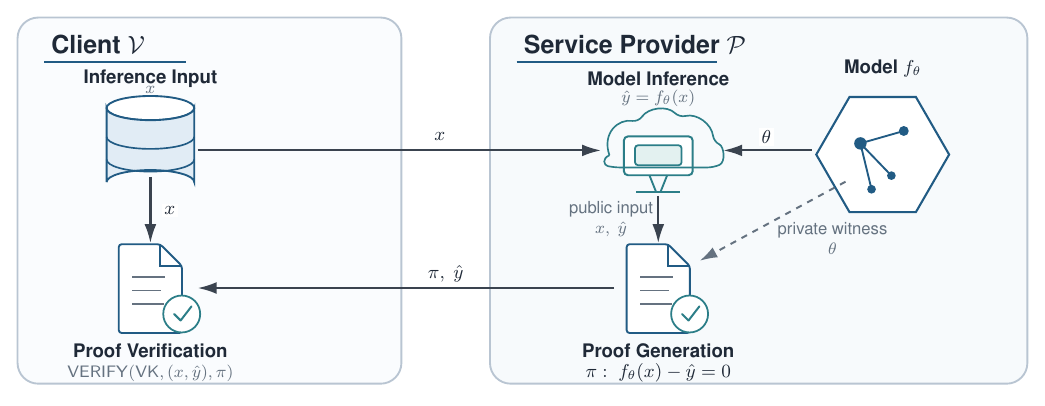}
	\caption{Illustrative workflow of ZKML-based verifiable inference. The client acts as the verifier \(\mathcal{V}\) and provides the inference input \(x\), whereas the ML service provider acts as the prover \(\mathcal{P}\) and evaluates the private model \(f_{\theta}\) to produce the prediction \(\hat{y}\). As shown in the figure, \((x,\hat{y})\) forms the public input to verification, while the model parameters \(\theta\) remain in the private witness. The prover returns \((\hat{y},\pi)\), and the verifier checks that the claimed output is consistent with the model execution without learning \(\theta\).}
	\label{f1}
\end{figure*}

\subsection{Illustrative ZKML Inference Workflow} Fig.~\ref{f1} presents an illustrative ZKML workflow for verifiable inference. In this setting, the ML service provider acts as the prover \(\mathcal{P}\), and the client acts as the verifier \(\mathcal{V}\). Let the client provide an inference input \(x\), and let the provider hold an ML model \(f_{\theta}\) with parameters \(\theta\). The provider computes the prediction \(\hat{y}=f_{\theta}(x)\) and then generates a proof \(\pi\) attesting that the returned output is consistent with the model execution. One convenient way to express the statement is through the relation \(F(x,\hat{y},\theta)=f_{\theta}(x)-\hat{y}=0\), where \((x,\hat{y})\) is treated as the public input and \(\theta\) is embedded in the private witness. The client finally verifies \(\pi\) and accepts \(\hat{y}\) only if the proof is valid. In this way, the client gains confidence that the reported prediction is genuine, while the provider preserves the confidentiality of its model parameters and internal computation process.

\subsection{Positioning Against Existing Surveys} Interest in ZKML has grown rapidly in both academia and industry, and the literature now spans circuit design, proof-system engineering, framework construction, and application-specific optimization. However, survey coverage of the area is still limited and fragmented. Existing surveys usually focus on only one slice of the field. Modulus Labs~\cite{moduluslabs2023cost} benchmarked practical ZKP systems for ML inference and highlighted proof-time and memory trade-offs. Sathe \emph{et al.}~\cite{sathe2023state} summarized several representative early ZKML systems but covered only a relatively small set of works. Xing \emph{et al.}~\cite{xing2023zero} focused on decentralized ML in communication networks and covered the literature up to June 2023. The open-source review in~\cite{zkp_framework} examined practical ZKP frameworks and development environments, but its main emphasis is framework usability rather than the broader ZKML research landscape. Therefore, the field still lacks an up-to-date survey that simultaneously covers algorithmic categories, technical evolution, implementation bottlenecks, commercial deployment, and future hybrid cryptographic directions. Table~\ref{tab:research_dimensions} compares these existing efforts with our survey.

\begin{table*}[t]
\centering

\caption{Comparison of existing ZKML surveys and this survey.}
\label{tab:research_dimensions}
\renewcommand\arraystretch{1.4}
\setlength{\tabcolsep}{5pt}
\begin{tabularx}{\textwidth}{@{}>{\raggedright\arraybackslash}p{4.1cm} >{\centering\arraybackslash}p{1.1cm} >{\raggedright\arraybackslash}X@{}}
  \toprule
  \textbf{Aspect} & \textbf{Ref.} & \textbf{Main Contributions} \\ 
  \midrule
  Benchmarking practical ZKP systems for ML inference & \cite{moduluslabs2023cost} &
  - Benchmarking six representative ZKP systems for ML inference; \par
  - Highlighting proof-time and memory trade-offs across different constructions; \par
  - Discussing practical deployment challenges through real-world case studies. \\
  \addlinespace[0.8em]
  Early survey of representative ZKML systems & \cite{sathe2023state} &
  - Summarizing early representative systems such as zkCNN, ezDPS, Xing, and Mystique; \par
  - Discussing CNN-oriented verification optimizations and implementation ideas. \\
  \addlinespace[0.8em]
  ZKP-based decentralized ML in communication networks & \cite{xing2023zero} &
  - Formalizing a four-algorithm ZKP-VML model; \par
  - Proposing distributed verification architectures (1-to-1, 1-to-$m$, and $m$-to-1); \par
  - Analyzing attack surfaces and defense mechanisms in networked settings. \\
  \addlinespace[0.8em]
  Assessment of open-source ZKP frameworks & \cite{zkp_framework} &
  - Comparing open-source ZKP frameworks, including constructions, environments, and benchmarks; \par
  - Evaluating scalability, runtime, proof size, and usability; \par
  - Providing practical recommendations for framework selection. \\
  \addlinespace[0.8em]
  \multicolumn{2}{@{}c}{Our work} &
  - Covering ZKP-based verifiable training, testing, and inference in a unified framework; \par
  - Reviewing the literature from 2017 to 2025 and synthesizing the technical evolution of the field; \par
  - Examining commercial applications and hybrid cryptographic frameworks for practical deployment; \par
  - Identifying open challenges and future directions in efficiency, generality, privacy, and scalability. \\
  \bottomrule
\end{tabularx}
\end{table*}

\subsection{Contributions} To address the above gap, we conducted a systematic survey of ZKML studies published from June 2017 to August 2025. We collected academic papers from major scientific databases, including ACM Digital Library, IEEE Xplore, Elsevier, MDPI, and SpringerLink. Since ZKML has also been actively explored in industry, we additionally reviewed white papers and technical blogs. Using keyword combinations related to zero-knowledge proofs and machine learning, we constructed a literature corpus that captures both foundational research and recent implementation-oriented developments. Based on this corpus, the main contributions of this survey are as follows:
\begin{itemize}
    \item \textbf{Comprehensive Research Landscape}: We provide a structured examination of ZKML research, organizing and analyzing 27 representative studies from June 2017 to August 2025.
    
    \item \textbf{Technical Evolution and Implementation Insights}: We categorize the major technical improvements in ZKML, especially those targeting efficiency and generality, and explain the design trade-offs behind these advances.
    
    \item \textbf{Bridge from Research to Practice}: We review commercial applications and practical frameworks for ZKML, highlighting how the field is evolving from conceptual prototypes toward deployable systems.
    
    \item \textbf{Open Problems and Future Directions}: We identify key technical bottlenecks and promising research directions, including computational optimization, model support, privacy enhancement, and scenario expansion.
\end{itemize}

The remainder of this paper is organized as follows. Section~\ref{s:bg} reviews the background of machine learning, zero-knowledge proofs, verifiable machine learning, and related security techniques. Section~\ref{s:3} surveys representative ZKML studies and their technical evolution. Section~\ref{s:4} discusses commercial applications of ZKML. Section~\ref{s:Hyb} examines hybrid cryptographic frameworks for verifiable ML. Section~\ref{s:5} outlines future directions for ZKML, and Section~\ref{s:6} concludes the paper. Table~\ref{tab1} lists the main abbreviations used throughout this paper.

\begin{table}[t]
    \centering
    \renewcommand{\arraystretch}{1.3}
    \setlength{\tabcolsep}{10pt}
    \caption{Main abbreviations used throughout this paper.}

    \begin{tabular}{@{}ll@{}}
        \toprule
        \textbf{Abbreviation} & \textbf{Full Name} \\ \midrule
        AI & Artificial Intelligence \\
        ML & Machine Learning \\
        MLaaS & Machine Learning as a Service \\
        ZKP & Zero-Knowledge Proof \\
        zk-SNARK & \makecell[l]{Zero-Knowledge Succinct\\ Non-interactive Argument of Knowledge} \\
        R1CS & Rank-1 Constraint System \\
        MLP & Multilayer Perceptron \\
        SVM & Support Vector Machine \\
        CRS & Common Reference String \\
        ROM & Random Oracle Model \\
        DP & Differential Privacy \\
        FL & Federated Learning \\
        TEE & Trusted Execution Environment \\
        SMC & Secure Multiparty Computation \\
        NN & Neural Network \\
        LogR & Logistic Regression \\
        LR & Linear Regression \\
        DWT & Discrete Wavelet Transformation \\
        PCA & Principal Component Analysis \\
        DT & Decision Tree \\
        CNN & Convolutional Neural Network \\
        LLM & Large Language Model \\
        QAP & Quadratic Arithmetic Problem \\
        QPP & Quadratic Polynomial Problem \\
        QMP & Quadratic Matrix Problem \\
        \bottomrule
    \end{tabular}
    \label{tab1}
\vspace{-15pt}
\end{table}

\section{Background}\label{s:bg}

This section reviews the technical foundations needed for the remainder of the survey. We first introduce the basic notation of machine learning and zero-knowledge proofs, then define the three main categories of verifiable machine learning, and finally compare ZKP with other representative security techniques used in trustworthy ML systems.

\subsection{Machine Learning}\label{ss:2a}

Machine learning (ML)~\cite{jordan2015machine} is a subfield of AI that learns predictive or decision-making functions from data and applies them to tasks such as classification, regression, pattern recognition, and control~\cite{janiesch2021machine}. Major paradigms include supervised learning, unsupervised learning, reinforcement learning, and self-supervised learning. Since most existing ZKML systems focus on proving the correctness of supervised training, testing, or inference, we adopt supervised learning as the default setting in this survey.

In supervised learning, let \(\mathcal{X}\) and \(\mathcal{Y}\) denote the input space and label space, respectively, and let \(g:\mathcal{X}\rightarrow\mathcal{Y}\) denote the target mapping. We represent an ML model by \(f_{\theta}:\mathcal{X}\rightarrow\mathcal{Y}\), where \(\theta\) denotes its trainable parameters. The model is trained on a dataset \(\mathcal{D}_{\mathrm{train}}=\{(x_i,y_i)\}_{i=1}^{n}\), where \(x_i\in\mathcal{X}\) is an input sample and \(y_i\in\mathcal{Y}\) is its corresponding label. The training procedure optimizes the model parameters by minimizing the empirical loss over \(\mathcal{D}_{\mathrm{train}}\), namely, \(\theta=\arg\min_{\theta'}\sum_{i=1}^{n}\ell\bigl(f_{\theta'}(x_i),y_i\bigr)\), where \(\ell(\cdot,\cdot)\) denotes the loss function. After training, for a new input \(x\in\mathcal{X}\), the model predicts the output \(\hat{y}=f_{\theta}(x)\). The objective is for the learned model \(f_{\theta}\) to approximate the target mapping \(g\), so that the prediction \(\hat{y}\) is as close as possible to the true label \(y\).

Various models can be used to represent the mapping from input features to output labels, including linear regression, decision trees, support vector machines, and deep neural networks. Each model has distinct characteristics and is suited to specific scenarios. Given the popularity and effectiveness of deep neural networks in modern applications, as well as the fact that nearly all existing work on ZKML focuses on deep neural networks, we adopt deep neural networks as the default ML model in this survey unless otherwise specified. Accordingly, \(\theta\) denotes the trainable parameters of the network, including weights and biases.

\subsection{Zero-Knowledge Proof}\label{ss:2b}
Zero-knowledge proof (ZKP) is a cryptographic technique that enables a prover \(\mathcal{P}\) to convince a verifier \(\mathcal{V}\) that a statement is true without revealing any additional information beyond the validity of that statement. Representative ZKP families include zero-knowledge succinct non-interactive arguments of knowledge (zk-SNARKs)~\cite{sasson2014zerocash}, zero-knowledge scalable transparent arguments of knowledge (zk-STARKs)~\cite{ben2019scalable}, and related constructions~\cite{canetti2000resettable}.


For many practical ZKP systems, the statement to be proven is encoded as an arithmetic circuit composed of addition, subtraction, multiplication, and division operations.\footnote{zk-SNARK systems are typically associated with arithmetic-circuit representations; zk-STARK can also support arithmetic-circuit representations, but in practice it often uses the more efficient arithmetic intermediate representation (AIR)~\cite{ben2018scalable}.} An \(\mathbb{F}\)-arithmetic circuit is a circuit in which all inputs and outputs are elements of a field \(\mathbb{F}\). Consider an \(\mathbb{F}\)-arithmetic circuit \(C\) with public input \(x \in \mathbb{F}^n\), witness \(w \in \mathbb{F}^h\), and output \(C(x,w) \in \mathbb{F}^l\), where \(n\), \(h\), and \(l\) denote the dimensions of the public input, witness, and output, respectively. The corresponding arithmetic-circuit satisfiability relation is \(\mathcal{R}_C = \{(x,w) \in \mathbb{F}^n \times \mathbb{F}^h : C(x,w) = 0^l\}\), and its induced language is \(\mathcal{L}_C = \{x \in \mathbb{F}^n : \exists w \in \mathbb{F}^h \text{ s.t. } C(x,w) = 0^l\}\). A ZKP system then consists of the following three algorithmic components~\cite{sasson2014zerocash}:

\begin{itemize}
    \item \((\mathsf{PK}, \mathsf{VK}) \leftarrow \mathsf{KEYGEN}(1^\lambda, C)\) is the key-generation algorithm that outputs the proving key \(\mathsf{PK}\) and the verification key \(\mathsf{VK}\) using the security parameter \(\lambda\) and the arithmetic circuit \(C\).
    \item \(\pi \leftarrow \mathsf{PROVE}(\mathsf{PK}, x, w)\) is the proof-generation algorithm that outputs a proof \(\pi\) given the proving key \(\mathsf{PK}\), the public input \(x\), and the witness \(w\).
    \item \(1/0 \leftarrow \mathsf{VERIFY}(\mathsf{VK}, x, \pi)\) is the proof-verification algorithm that outputs an accept/reject decision using \(\mathsf{VK}\), \(x\), and \(\pi\) as input.
\end{itemize}
The proving key \(\mathsf{PK}\) and the verification key \(\mathsf{VK}\) generated by \(\mathsf{KEYGEN}\) are public parameters. The prover \(\mathcal{P}\) executes \(\mathsf{PROVE}\), and the verifier \(\mathcal{V}\) executes \(\mathsf{VERIFY}\). The witness \(w\) is the secret held by \(\mathcal{P}\) that should remain hidden while still being proven known.

Different ZKP systems are suited to different circuit representations and computational structures. For example, zk-SNARKs are particularly effective for arithmetic circuits expressed in Rank-1 Constraint System (R1CS) form, whereas systems based on the sum-check~\cite{lund1992algebraic} and GKR~\cite{thaler2015note} protocols are often more suitable for layered computations. This distinction is highly relevant to ML, because deep neural networks naturally exhibit layered structures. Consequently, many ZKML systems for neural-network verification build on sum-check- or GKR-style proving frameworks to improve efficiency.

For ZKML, two properties of ZKP are particularly important. First, \(\mathsf{VERIFY}\) is usually much cheaper to execute than \(\mathsf{PROVE}\), which allows a resource-constrained client to validate an expensive outsourced ML computation efficiently. Second, \(\mathsf{VERIFY}\) does not require access to the private witness, which means that sensitive information such as model parameters, training data, or internal intermediate states can remain hidden throughout the verification process.

\subsection{Verifiable Machine Learning}\label{ss:2c}
As ML models become larger and more expensive, outsourcing training, testing, and inference to external platforms has become increasingly common. In such settings, a client typically lacks the resources required to repeat the entire computation locally and therefore needs a mechanism to verify the integrity of the outsourced result. Verifiable machine learning studies how a service provider can furnish convincing evidence that an ML task has been executed according to a declared dataset, model, and evaluation procedure.

In the ZKML setting, the service provider typically acts as the prover \(\mathcal{P}\), and the client acts as the verifier \(\mathcal{V}\). Depending on the stage of the ML pipeline being certified, verifiable machine learning can be divided into three primary categories: \emph{Verifiable Training}, \emph{Verifiable Testing}, and \emph{Verifiable Inference}.

\begin{itemize}
    \item {\bf \emph{Verifiable Training}} certifies that the updated model parameters \(\theta'\) are obtained by running the declared training procedure on the committed training data under the agreed hyperparameters and architecture. This category is relevant when a client outsources model training but still needs evidence that the delivered model is the genuine result of the specified optimization process rather than a fabricated or improperly trained artifact.

    \item {\bf \emph{Verifiable Testing}} certifies that a reported evaluation metric, such as accuracy or F1 score, is computed from the declared test dataset and the committed model under a specified evaluation protocol. Its goal is to prevent misleading claims about model quality, for example when a provider reports inflated performance that does not correspond to the agreed test setting.

    \item {\bf \emph{Verifiable Inference}} certifies that the returned prediction \(\hat{y}\) is exactly the output produced by the specified model on the declared input \(x\). This category is especially important in MLaaS scenarios where the client wants assurance that the inference result is genuine, while the provider wants to keep the model parameters confidential.
\end{itemize}

\begin{table*}[t]
\caption{Comparison of representative security techniques for verifiable machine learning.}
\centering
\renewcommand{\arraystretch}{1.2}
\setlength{\tabcolsep}{4pt}
\resizebox{\textwidth}{!}{%
\begin{tabular}{@{}lllll@{}}
\toprule
\makecell[l]{\textbf{Technique}} & \makecell[l]{\textbf{Security} \\ \textbf{Basis}} & \makecell[l]{\textbf{Primary Security} \\ \textbf{Property}} & \makecell[l]{\textbf{Runtime} \\ \textbf{Overhead}} & \makecell[l]{\textbf{Verification} \\ \textbf{Capability}} \\ \midrule
DP & Statistical privacy model & Record-level privacy via randomized noise & Low & None \\ 
HE & Cryptographic hardness assumptions & Confidential computation on encrypted data & Very high & None \\ 
FL & Trust in protocol participants / coordinator & Data locality and collaborative training & Low to moderate & None \\ 
TEE & Trust in hardware vendor and platform & Hardware-backed execution isolation & Low & Limited (attestation) \\ 
SMC & Cryptographic assumptions & Multi-party input privacy & High & Limited \\ 
ZKP & Cryptographic assumptions & Privacy plus verifiable computation & Moderate to high & Strong \\ 
\bottomrule
\end{tabular}
}
\label{tabCom}
\end{table*}

\subsection{Comparison of Security Techniques for Verifiable Machine Learning}\label{ss:2d}

Several security techniques, including Differential Privacy (DP), Homomorphic Encryption (HE), Federated Learning (FL), Trusted Execution Environments (TEE), and Secure Multiparty Computation (SMC), are important for privacy-preserving and trustworthy ML. However, they target different threat models and provide different guarantees. In particular, none of them, when used alone, offers the same combination of privacy preservation and succinct public verifiability as ZKP. Table~\ref{tabCom} summarizes the comparison, and the main distinctions are outlined below.

DP is a rigorous privacy framework that protects individual records by injecting calibrated random noise~\cite{dwork2006differential,dwork2008differential}. In ML, DP is widely used to reduce the leakage risk of training data. However, the added noise may reduce model utility, and DP does not provide any mechanism for proving that the claimed privacy budget or computational procedure was actually enforced.

HE enables direct computation on encrypted data~\cite{yi2014homomorphic}. This makes HE highly attractive for confidential inference and secure outsourced computation. Its main limitation is efficiency: ciphertext expansion, noise management, and expensive homomorphic operations often introduce substantial computational and storage overhead. Moreover, HE alone does not let a client verify that the service provider executed the encrypted computation correctly.

FL keeps raw data local and supports collaborative model training across multiple participants~\cite{konevcny2016federated, rieke2020future}. This data-locality property can improve privacy and regulatory compliance. Nevertheless, FL does not natively prevent malicious participants from submitting poisoned or low-quality updates, and it offers no inherent cryptographic proof that the aggregation or local training steps were executed correctly.

TEE provides hardware-backed isolation for storing data and executing code securely~\cite{sabt2015trusted}. Compared with purely cryptographic techniques, TEE is usually efficient and can support complex programs with low runtime overhead. Its limitations stem from its trust model: users must trust the hardware vendor and platform implementation, and the resulting security can be weakened by side-channel attacks, limited enclave memory, and implementation vulnerabilities. Although remote attestation offers some assurance about the execution environment, it is not equivalent to a succinct public proof of each computation step.

SMC enables multiple parties to jointly compute a function while keeping their private inputs hidden~\cite{zhao2019secure}. It offers strong privacy guarantees and is well suited to collaborative settings. However, SMC protocols often incur high communication overhead, multiple interaction rounds, and significant implementation complexity, which makes large-scale ML deployment challenging. In addition, the resulting guarantees are usually protocol-specific rather than succinctly verifiable by an arbitrary third party.

Overall, these techniques should be viewed as complementary rather than mutually exclusive. ZKP is distinctive because it directly addresses computational verifiability while still protecting private witnesses. This makes it a natural foundation for verifiable ML, and it also explains why recent systems increasingly combine ZKP with HE, FL, DP, or TEE to obtain stronger end-to-end guarantees. Section~\ref{s:Hyb} revisits these hybrid designs in more detail.

\begin{figure*}[t]
	\centering
	\includegraphics[width=\linewidth]{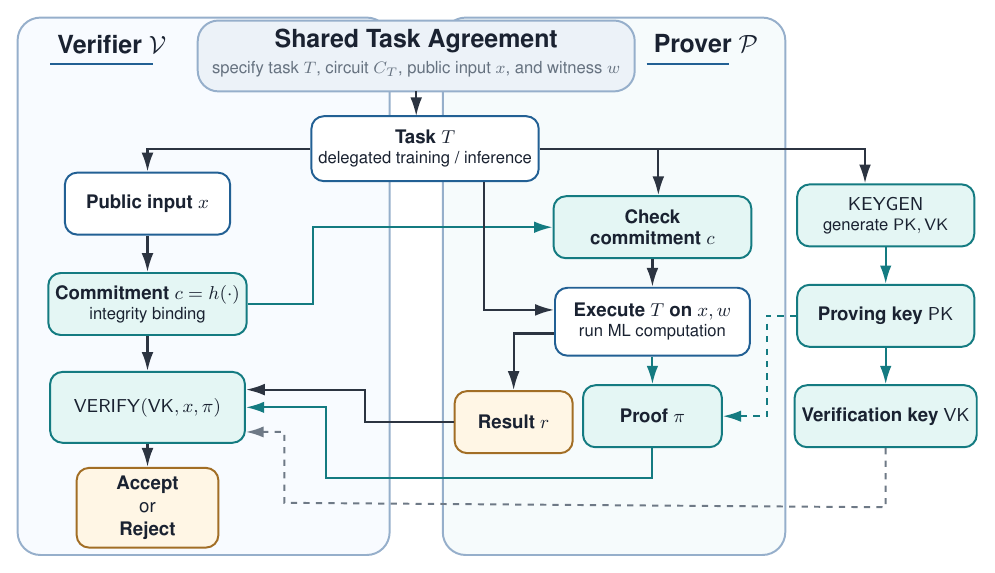}
	\caption{Typical workflow of a ZKML protocol. The verifier \(\mathcal{V}\) and prover \(\mathcal{P}\) first establish a shared task agreement specifying the task \(T\), the circuit \(C_T\), the public input \(x\), and the witness \(w\). The verifier commits to the relevant information through a digest \(c=h(\cdot)\), while the setup procedure generates \((\mathsf{PK},\mathsf{VK})\). The prover then checks the commitment, executes the delegated ML task on \((x,w)\), obtains the result \(r\), and generates a proof \(\pi\). Finally, the verifier runs \(\mathsf{VERIFY}(\mathsf{VK},x,\pi)\) and outputs an accept/reject decision.}
	\label{f3}
\end{figure*}

\section{ZKML Research Landscape} \label{s:3}

This section reviews the research landscape of ZKML. We first outline the general workflow and design logic of ZKML systems, then survey representative studies across verifiable training, testing, and inference, and finally summarize the main implementation challenges together with the technical improvements proposed to address them.

\subsection{ZKML Overview}\label{ss:3a}
In ZKML, there are two key participants: the ML service provider acts as the prover in the ZKP system, denoted by \(\mathcal{P}\), and the ML client acts as the verifier, denoted by \(\mathcal{V}\). Because \(\mathcal{V}\) may have limited computational resources or limited access to the required data and models, it delegates computationally intensive ML tasks, such as training, testing, or inference, to \(\mathcal{P}\). Accordingly, most of the ML computation is carried out by \(\mathcal{P}\), while \(\mathcal{V}\) focuses on checking the correctness and integrity of the returned result.

Depending on the task setting and privacy requirements, \(\mathcal{P}\) and \(\mathcal{V}\) may possess different datasets, models, and verification objectives. Since both parties operate in a trustless environment, a dishonest prover may deviate from the agreed workflow, for example by using incorrect data, skipping part of the computation, or fabricating a seemingly valid result. ZKML addresses this problem by attaching cryptographic verifiability to the ML computation.

The typical workflow of ZKML is illustrated in Fig.~\ref{f3}. First, the prover \(\mathcal{P}\) and the verifier \(\mathcal{V}\) agree on a task instance \(T\), the corresponding arithmetic circuit \(C_T\), and the public information to be exposed during verification. The verifier \(\mathcal{V}\) commits to the relevant data or model state through a digest \(c=h(\cdot)\) and sends this commitment to \(\mathcal{P}\). After the setup phase generates \((\mathsf{PK},\mathsf{VK}) \leftarrow \mathsf{KEYGEN}(1^\lambda,C_T)\), the prover \(\mathcal{P}\) executes the ML task, obtains the result \(r\), and produces a proof \(\pi \leftarrow \mathsf{PROVE}(\mathsf{PK},x,w)\), where \(x\) and \(w\) denote the public input and private witness, respectively. Finally, the verifier \(\mathcal{V}\) checks the claimed result together with \(\pi\) by running \(\mathsf{VERIFY}(\mathsf{VK},x,\pi)\). If verification succeeds, \(\mathcal{V}\) accepts that \(\mathcal{P}\) has correctly executed \(T\); otherwise, it rejects the result.

In a ZKP system, the statement to be proven (\emph{e.g.}, verifying that a computation was performed correctly or that a secret satisfies certain conditions) is encoded as a computation. This computation is typically represented as an arithmetic circuit, which processes public inputs and private witnesses through a sequence of arithmetic operations to generate an output. In the following discussion, we use \(h(\cdot)\) to denote a hash-based commitment, \(\mathcal{D}_{\mathrm{train}}=(X_{\mathrm{train}},Y_{\mathrm{train}})\) and \(\mathcal{D}_{\mathrm{test}}=(X_{\mathrm{test}},Y_{\mathrm{test}})\) to denote the training and testing datasets, \(\theta\) to denote model parameters, \(\mathsf{PK}\) and \(\mathsf{VK}\) to denote the proving and verification keys, and \(\pi\) to denote the generated proof. The conceptual diagrams in this paper highlight three primary components: circuit input, circuit logic, and circuit output. Their configurations vary across the three categories of ZKP-based verifiable machine learning, as described below.

\subsubsection{\bf ZKP-based Verifiable Training} 
\begin{figure*}[!t]
    \centering
    \includegraphics[width=\linewidth]{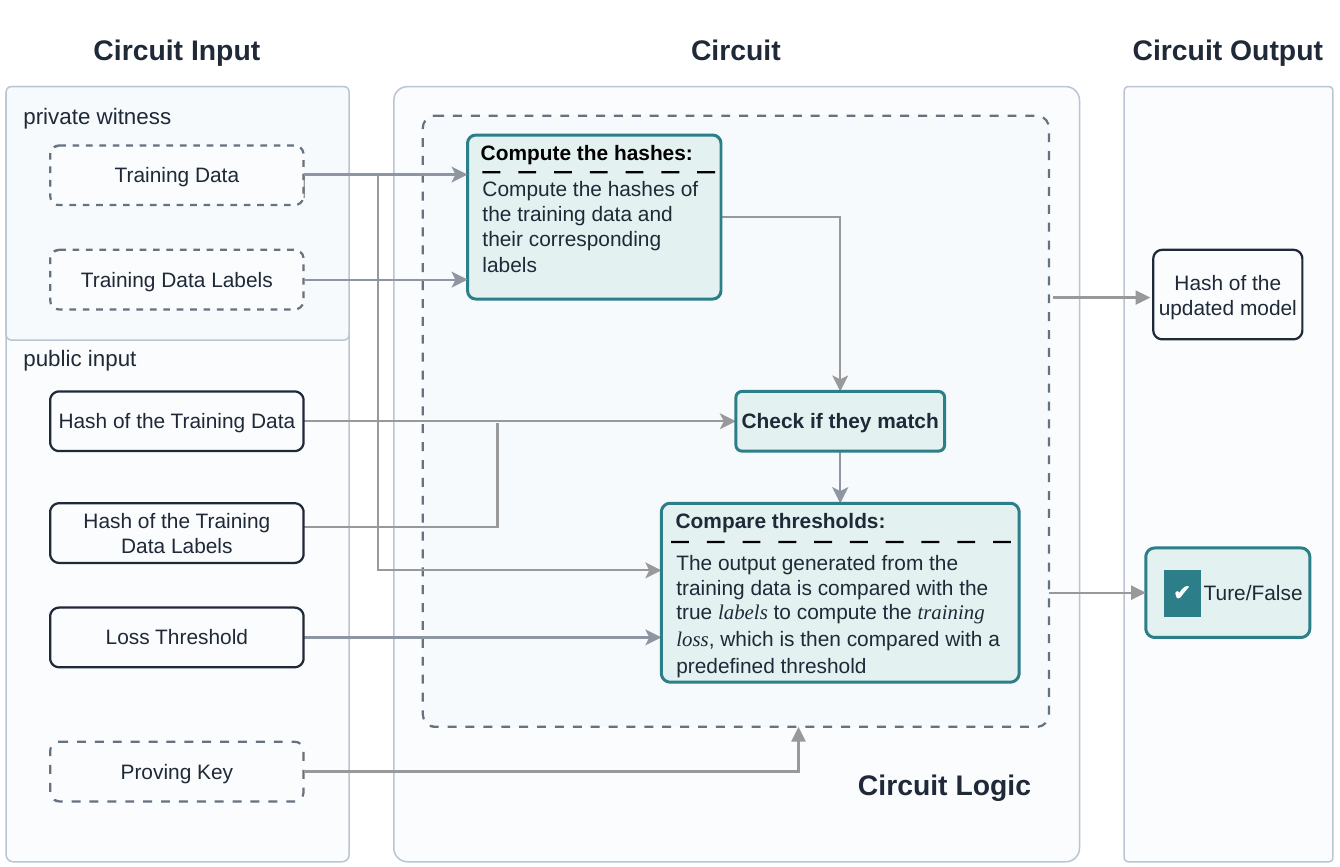}
    \caption{Conceptual arithmetic-circuit design for ZKP-based verifiable training. The figure separates the computation into circuit input, circuit logic, and circuit output. The public inputs include the hashes of the training data and labels, the loss threshold, and the proving key, whereas the private witness contains the underlying training data and labels. Inside the circuit, the hashes are recomputed and checked for consistency, after which the generated training output is compared with the ground-truth labels to evaluate whether the declared loss condition is satisfied. The circuit output shown in the figure consists of the hash of the updated model together with a true/false validity outcome.}
    \label{f5}
\end{figure*}

Fig.~\ref{f5} presents the conceptual diagram of the arithmetic circuits used in ZKP-based verifiable training. Below, we detail the components of this diagram. 

\begin{itemize}

\item {Circuit Input}: The circuit takes both a private witness and a public input. The witness contains the training features and labels, denoted by \((X_{\mathrm{train}},Y_{\mathrm{train}})\). The public input contains the commitments \(h(X_{\mathrm{train}})\) and \(h(Y_{\mathrm{train}})\), the target loss threshold \(\tau_{\mathrm{loss}}\), and the proving key \(\mathsf{PK}\).\footnote{The hashes used in the inputs serve as commitments to the corresponding data. Alternatively, other cryptographic commitment schemes compatible with ZKP can also be utilized.}

\item {Circuit Logic}: The circuit first recomputes \(h(X_{\mathrm{train}})\) and \(h(Y_{\mathrm{train}})\) from the witness and checks them against the public commitments. If the commitments are consistent, it evaluates the model on \(X_{\mathrm{train}}\), computes the training loss against \(Y_{\mathrm{train}}\), and compares the resulting loss with \(\tau_{\mathrm{loss}}\). If the loss remains above the threshold, the circuit computes gradients and updates the model parameters; otherwise, the training process halts. The entire computation is coupled with \(\mathsf{PK}\) to generate a zero-knowledge proof \(\pi\), thereby ensuring integrity without revealing the private training data.

\item {Circuit Output}: The output includes the commitment \(h(\theta')\) of the updated model parameters \(\theta'\) and the generated proof \(\pi\). 

\end{itemize}

This circuit design enables \(\mathcal{V}\) to verify that the updated model commitment \(h(\theta')\) is obtained from the declared training workflow and the committed training dataset, without revealing the private training data or model parameters. Consequently, it ensures the transparency and trustworthiness of the training process while preserving data and model privacy.

\subsubsection{\bf ZKP-based Verifiable Testing} 
\begin{figure*}[!t]
	\centering
	\includegraphics[width=\linewidth]{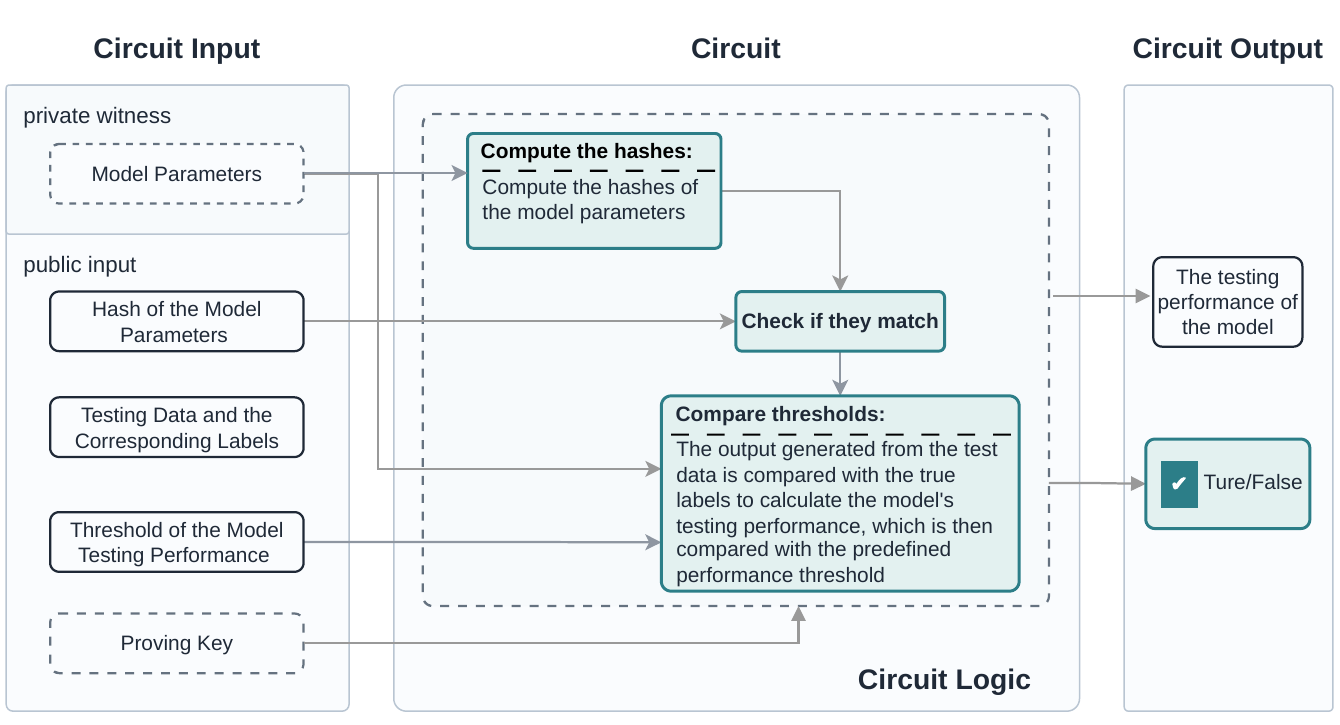}
	\caption{Conceptual arithmetic-circuit design for ZKP-based verifiable testing. The figure shows a three-part view consisting of circuit input, circuit logic, and circuit output. The public input contains the hash of the model parameters, the testing data and corresponding labels, the performance threshold, and the proving key, while the private witness contains the model parameters themselves. The circuit recomputes the parameter hash, checks consistency with the public commitment, evaluates the model on the declared test set, and compares the resulting testing performance against the stated threshold. The circuit output shown here is the testing performance of the model together with a true/false validity outcome.}
	\label{f6}
\end{figure*}

Fig.~\ref{f6} presents the conceptual diagram of the arithmetic circuits used in ZKP-based verifiable testing. Below, we detail the components of this diagram.

\begin{itemize}

\item {Circuit Input}: The private witness contains the model parameters \(\theta\). The public input contains the testing dataset \(\mathcal{D}_{\mathrm{test}}=(X_{\mathrm{test}},Y_{\mathrm{test}})\), the parameter commitment \(h(\theta)\), the performance threshold \(\tau_{\mathrm{perf}}\), and the proving key \(\mathsf{PK}\).

\item {Circuit Logic}: The circuit first recomputes \(h(\theta)\) from the witness and compares it with the public commitment. If the commitments match, it evaluates the model on \(X_{\mathrm{test}}\), computes the testing metric \(m_{\mathrm{test}}\) against \(Y_{\mathrm{test}}\), and compares the resulting score with \(\tau_{\mathrm{perf}}\). If the performance requirement is satisfied, the circuit uses \(\mathsf{PK}\) to generate the proof \(\pi\).

\item {Circuit Output}: The output includes the testing metric \(m_{\mathrm{test}}\) and the generated proof \(\pi\).

\end{itemize}

This circuit design enables \(\mathcal{V}\) to verify that the reported testing metric \(m_{\mathrm{test}}\) is computed from the committed model \(\theta\) and the declared testing dataset \(\mathcal{D}_{\mathrm{test}}\), while keeping the model parameters hidden.

\subsubsection{\bf ZKP-based Verifiable Inference} 
\begin{figure*}[!t]
	\centering
	\includegraphics[width=\linewidth]{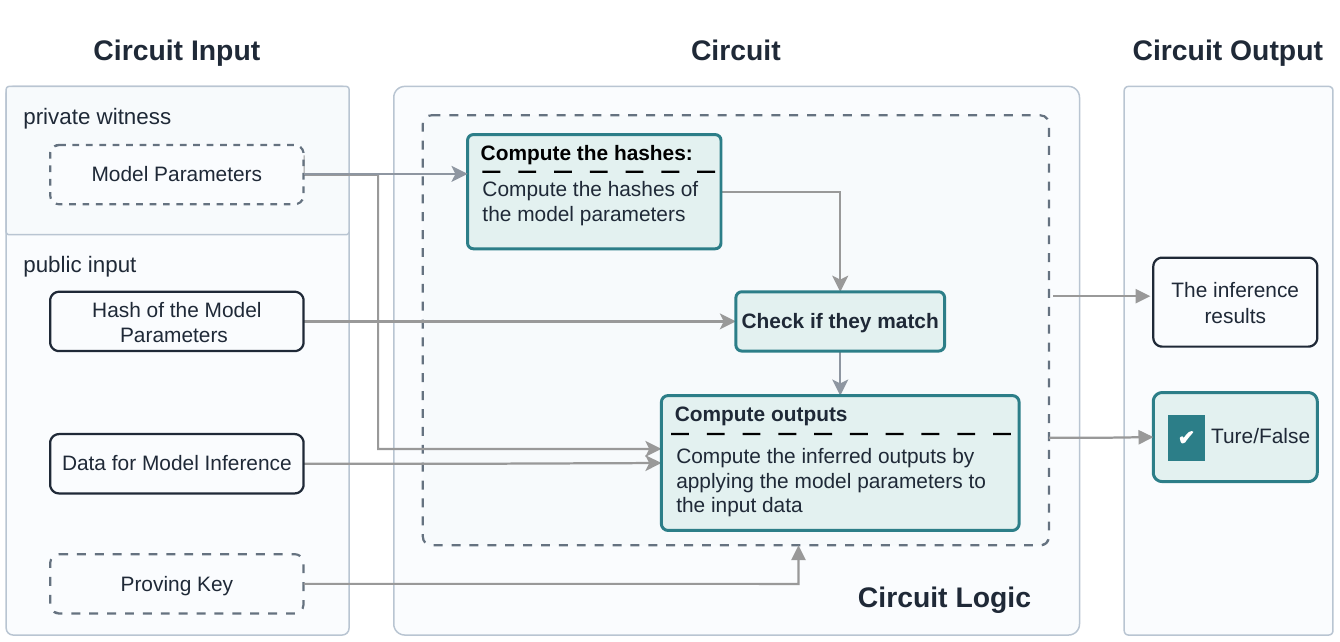}
	\caption{Conceptual arithmetic-circuit design for ZKP-based verifiable inference. The figure shows that the public input contains the hash of the model parameters, the data for model inference, and the proving key, whereas the private witness contains the model parameters themselves. Inside the circuit, the parameter hash is recomputed and checked against the public commitment before the model is applied to the input data to compute the inference result. The circuit output shown in the figure is the inference result together with a true/false validity outcome, allowing correctness to be certified without revealing the private model parameters.}
	\label{f7}
	\vspace{1cm}
\end{figure*}

Fig.~\ref{f7} presents the conceptual diagram of the arithmetic circuits used in ZKP-based verifiable inference. Below, we detail the components of this diagram.

\begin{itemize}

\item {Circuit Input}: The inputs to the circuit in ZKP-based verifiable inference include both a private witness and a public input. The witness contains the model parameters \(\theta\). The public input includes the parameter commitment \(h(\theta)\), the inference input \(x\), and the proving key \(\mathsf{PK}\).

\item {Circuit Logic}: The circuit first recomputes \(h(\theta)\) from the witness and compares it with the public commitment. If the commitments match, it evaluates the model on the input \(x\) to obtain the prediction \(\hat{y}=f_{\theta}(x)\). At the same time, \(\mathsf{PK}\) is used to generate a zero-knowledge proof \(\pi\) for the correctness of the inference computation. 

\item {Circuit Output}: The output consists of the prediction \(\hat{y}\) and the generated proof \(\pi\). 

\end{itemize}

This circuit design enables \(\mathcal{V}\) to verify that the prediction \(\hat{y}\) is generated by the committed model \(\theta\) on the declared input \(x\), without exposing the model parameters. Consequently, it preserves model privacy while maintaining the integrity and trustworthiness of the inference result.

Based on the conceptual diagrams of the arithmetic circuits, we can intuitively illustrate how the objectives of the three categories of ZKML are achieved using ZKP. In the following part of this section, we classify existing ZKML studies into these three categories and analyze them individually.

\subsection{Representative ZKML Studies} \label{ss:3b}

\begin{figure*}[t]
	\centering
	\includegraphics[width=\linewidth]{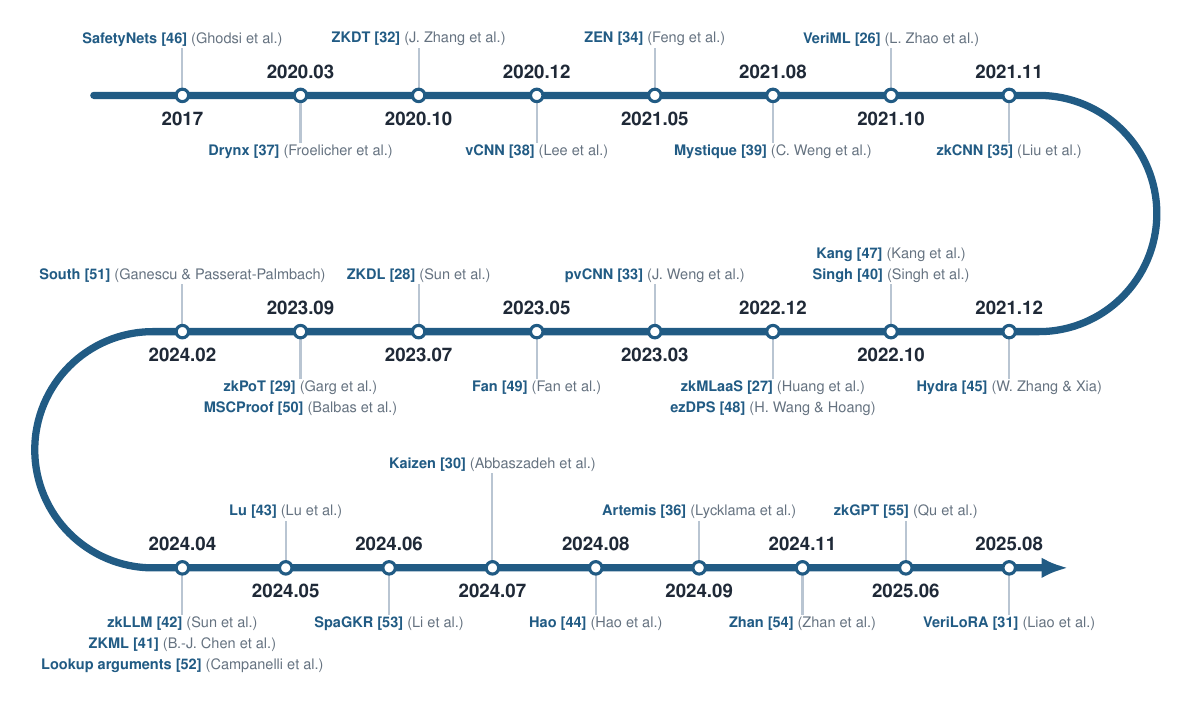}
	\caption{Timeline of representative ZKML studies. The figure highlights the evolution of the field from early proof-of-concept systems for neural-network inference and decision-tree verification to more recent work on verifiable training, commit-and-prove optimization, large-model inference, and LLM-oriented systems. It also illustrates how the research focus gradually shifted from small and highly specialized models toward broader system support, higher efficiency, and stronger deployment realism.}
	\label{f4}
\end{figure*}

Since the emergence of SafetyNets in 2017, ZKML has evolved from early proofs of concept into a diverse research landscape spanning training, testing, and inference verification. Fig.~\ref{f4} outlines the timeline of representative studies. For clarity, we organize these works by verification objective, namely verifiable training, verifiable testing, and verifiable inference. Their overall classification is summarized in Table~\ref{tab3}. To facilitate direct efficiency comparisons, we further report proving time, verification time, and proof size in three dedicated quantitative tables: Table~\ref{tab:training} for verifiable training, Table~\ref{tab:testing} for verifiable testing, and Table~\ref{tab:inference} for verifiable inference. The discussion below focuses on the core design ideas, implementation strategies, and application settings of representative systems, thereby clarifying how ZKML has progressed toward more trustworthy and practical machine learning services.

\begin{table*}[t]
    
    \caption{Quantitative comparison of representative ZKP-based verifiable training systems.}
    
    \centering
    \small    \resizebox{1\textwidth}{!}{
        \begin{tabular}{l L{2.5cm} L{2.7cm} ccc}
        \toprule
        \textbf{Work} & \textbf{Model(s)} & \textbf{Hardware / Platform} & \textbf{Proving Time} & \textbf{Verification Time} & \textbf{Proof Size} \\ 
        \midrule
        VeriML~\cite{zhao2021veriml} & Six models (e.g., NN) & Intel Core i5-4460S & $<20$s (e.g., 12s for NN, batch size 32) & $<2$s (e.g., 0.99s for NN) & N/A\\
        zkMLaaS~\cite{huang2022zkmlaas} & LogiR & 24 Intel Xeon E5-262, 8 V100 GPUs, 128GB RAM& 2.2s & 5ms & 24.1MB \\
        ZKDL~\cite{sun2023zkdl} & DNN & 12 CPU cores, Tesla A100 GPU, 128 GB RAM& 0.86s (1024 samples, batch size 64) & 0.19s & 0.033KB \\
        ZKPoT~\cite{garg2023experimenting} & LogiR & N2 GCP, 512GB RAM & 4208s (total)& 26.5s (total)& $<350$MB \\
        Kaizen~\cite{abbaszadeh2024zero} & VGG-11 (10M parameters) & 8x Xeon Platinum 8370, 512GB RAM & $<882$s (per sample, batch size 16) & $<0.130$s & $<1627$KB \\
        VeriLoRA~\cite{liao2025zklora} & LLM ($<$13B, e.g., LLaMA-2) & AMD EPYC 9654, Nvidia A100 (80GB), 500GB RAM & $<600$s (per sample, batch size 1) & $<4$s & N/A\\
         \bottomrule

    \end{tabular}}
    \label{tab:training}
\end{table*}

\begin{table*}[t]
    \caption{Quantitative comparison of representative ZKP-based verifiable testing systems.}
    
    \centering
    \small    \resizebox{1\textwidth}{!}{
    \begin{tabular}{l L{2.5cm} L{2.7cm} ccc}
        \toprule
        \textbf{Work} & \textbf{Model(s)} & \textbf{Hardware / Platform} & \textbf{Proving Time} & \textbf{Verification Time} & \textbf{Proof Size} \\
        \midrule
        ZKDT~\cite{zhang2020zero} & DT (1029 nodes, 23 levels, 5000 samples) & Intel Xeon Platinum 8124M & 250s (accuracy testing) & 15.6s & 287KB \\
        pvCNN~\cite{weng2023pvcnn} & CNN (e.g., LeNet-5) & Quad-core i5-7500, 48GB RAM& 1448.83s (ReLU matrix) & 54s (MNIST) & 351MB \\
        ZEN~\cite{feng2021zen} & NN (ShallowNet, LeNet) & 32-core Xeon 8280M, 256GB RAM& 147s to 4710s & 0.023s to 0.47s & 192 bytes \\
        zkCNN~\cite{liu2021zkcnn} & CNN & AMD EPYC 7R32, 128GB RAM & 88.3s & 59.3ms & 341KB \\
        Artemis~\cite{lycklama2024artemis} & ML (GPT-2)& 128 vCPUs, 1TB RAM & 200-240 min (GPT-2) & 14ms (GPT-2) & 15KB (GPT-2) \\ 
        \bottomrule
    \end{tabular}}
    \label{tab:testing}
\end{table*}

\begin{table*}[t]
    \caption{Quantitative comparison of representative ZKP-based verifiable inference systems.}
    
    \centering
    \small    \resizebox{1\textwidth}{!}{
    \begin{tabular}{l L{2.5cm} L{2.7cm} cc>{\centering\arraybackslash}p{2cm}}
        \toprule
        \textbf{Work} & \textbf{Model(s)} & \textbf{Hardware / Platform} & \textbf{Proving Time} & \textbf{Verification Time} & \textbf{Proof Size} \\
        \midrule
        Drynx~\cite{froelicher2020drynx} & Regression models & Intel Xeon E5-2680 v3 CPUs, 256GB RAM  & $<22$s (LogR, 600k records) & $<22$s & 56kB \\
        vCNN~\cite{lee2024vcnn} & CNNs (VGG16) & Quad-core Intel CPU i5 & 8 hours (VGG16) & 19.4s (VGG16) & N/A \\
        Mystique~\cite{weng2021mystique} & NN (ResNet-101, 42.5M parameters) & m5.2xlarge, 32GB RAM& 262s & N/A & 0.99GB \\
        Singh~\cite{singh2022zero} & DT & 8x Xeon, 32GB RAM & 170-200s & 0.4s & N/A \\
        ZKML~\cite{chen2024zkml} & ML (GPT-2) & 128 vCPUs, 1TB RAM & 3652s (GPT-2) & 18.7s (GPT-2) & 28KB (GPT-2) \\
        zkLLM~\cite{sun2024zkllm} & LLM ($<$13B, e.g., LLaMA-2) & Nvidia A100 (40GB), 124GB RAM& $<15$ min & $<3$s & $<10$KB \\
        Lu~\cite{lu2024efficient} & NN (GPT-2, 117M parameters) & Ryzen 3700X 8-core, 32GB RAM& 287.1s (GPT-2) & 63s (GPT-2) & 4.5GB (GPT-2) \\
        Hao~\cite{hao2024scalable} & LLMs & AWS c5.9xlarge & 2.1s (ReLU); 87s (Softmax) & N/A & 30MB (ReLU); 816MB (Softmax) \\
        \bottomrule
    \end{tabular}}
    \label{tab:inference}
\end{table*}

\subsubsection{\bf ZKP-based Verifiable Training}

In 2021, \cite{zhao2021veriml} proposed VeriML, whose core idea is to make the training process retrievable to achieve verifiability in machine learning training. VeriML pre-stores the inputs and outputs of several iterations during the training process and commits to them, allowing the prover to retrieve specified iterations upon the verifier's request and generate proofs of their computational processes. VeriML supports six typical machine learning models, including linear regression, logistic regression, neural networks, SVM, K-Means, and decision trees. Also, VeriML's computational and communication costs are justified through numerous experiments. Experimental results demonstrate that the communication overhead of VeriML is related to the number of iterations and the number of data samples, and the computational overhead of VeriML is dominated by the batch size of data samples and increases linearly with the batch size.

\cite{zhang2021hydra} introduced Hydra, a verifiable training protocol for neural networks built upon the GKR protocol. This system is specifically tailored for neural networks and builds upon the SafetyNets method---the first verifiable inference approach leveraging ZKP, which will be discussed later. In particular, Hydra uses SafetyNets to represent a neural network as an arithmetic circuit. To achieve verifiable training, Hydra introduces a subcircuit protocol and a  interactive quantization algorithm. The execution process of Hydra's subcircuit protocol can be summarized as follows: First, the large and deep circuit representing a neural network is partitioned into multiple smaller subcircuits by depth. The prover then applies the GKR protocol to generate proofs for each sub-circuit. Once a proof for a sub-circuit is generated, it is immediately transmitted to the verifier for verification. Finally, all proofs corresponding to the sub-circuits are aggregated into a single final proof for the entire circuit, which is sent to the verifier for verification. This pipelined approach enables the proof-and-verification process to begin as soon as sub-circuits are uploaded, eliminating the need to process the entire circuit at once. By overlapping proof generation and verification, Hydra significantly enhances verification efficiency. Hydra's interactive quantization algorithm begins by rounding the weights of the bottom layers of the neural network, freezing these quantized layers, and retraining the network. This process is then repeated layer by layer, progressively moving upward, until all layers have been quantized. This interactive quantization algorithm is executed for each part of the neural network corresponding to each sub-circuit in the pipelined proof-and-verification process.

\cite{huang2022zkmlaas} proposed zkMLaaS, a framework that employs a two-round challenge-response protocol and random sampling of model weights to generate proofs. This approach reduces the time cost of proof generation while ensuring the verifiable integrity of the training process. Before executing the challenge-response protocol, an honest third party (or the ML client) generates a key pair \((\mathsf{PK}, \mathsf{VK})\), distributing \(\mathsf{PK}\) to the ML service provider \(\mathcal{P}\) and \(\mathsf{VK}\) to the ML client \(\mathcal{V}\). In the first round of the challenge-response protocol, the ML service provider submits commitments for all intermediate weights updated during each training iteration and data sampling epoch to the ML client. In the second round, the ML client randomly selects a subset of these intermediate weights, and the service provider must generate corresponding proofs. The zkMLaaS framework leverages zk-SNARKs and commitment schemes to ensure security. The binding property of the commitment scheme guarantees that the ML service provider cannot produce two different sets of weights corresponding to the same commitment. Furthermore, since the public key is generated by the client or a trusted third party, the service provider is restricted to generating proofs only for the circuits specified by the client. This ensures that the intermediate weights are derived solely from the correct model and data.

zkDL is an efficient ZKML system designed for the verifiable training of deep neural networks~\cite{sun2023zkdl}. To address the non-arithmetic nature of ReLU activation functions in neural networks, zkDL introduces the zkReLU protocol, which enables efficient proof generation for computations involving ReLU. The zkReLU protocol leverages auxiliary inputs to transform the nonlinear computations of ReLU into an equivalent linear form, enabling efficient handling of forward and backward propagations through ReLU. To ensure the validity of these auxiliary inputs, zkReLU employs the Pedersen commitment scheme and incorporates an anchoring mechanism that links the arithmetic operations within each layer to the auxiliary inputs. This design not only preserves the tensor structure inherent in deep neural networks but also significantly reduces computational redundancy by reusing verified commitments. For circuit design in the training of deep neural networks, zkDL proposes FAC4DNN (Flat Arithmetic Circuit for Deep Neural Networks), which parallelizes the traditionally sequential execution of neural network layers and training iterations. This approach effectively flattens the circuit, reducing its depth by a factor of $O(N)$, where $N$ is the product of the neural network depth and the number of training iterations. zkDL implements a three-step parallel proof process for FAC4DNN: (1) parallel execution of the GKR protocol for each layer, (2) parallel proof of inter-layer arithmetic relationships, and (3) parallel verification of auxiliary inputs. This design ensures that the proof size grows only by $O(\log L)$, where $L$ is the product of the neural network depth. Compared to traditional sequential proof generation, zkReLU's parallel proof generation offers significant time efficiency. It shows in~\cite{sun2023zkdl} that the proof generation of zkDL for the entire forward and backward propagation of the neural network training process to be completed within tens of seconds for neural networks with 10 millions parameters.

\cite{garg2023experimenting} introduced Zero-Knowledge Proof of Training (zkPoT), a proof-of-training framework aimed at simple models such as logistic regression. Its key idea is to separate arithmetic and non-arithmetic parts of the training computation so that proof size scales with the number of training samples rather than the feature dimension. The design also streams data from auxiliary memory instead of requiring the full computation trace to remain in RAM, making it better aligned with the memory demands of training-oriented workloads.

\begin{table*}[t]
    \caption{Classification of representative ZKML systems by verification objective.}
    \centering
    \small    \resizebox{1\textwidth}{!}{
    \begin{tabular}{l c l c c c}
        \toprule
        \textbf{Work} & \textbf{Date} & \textbf{Model / Scenario} & \textbf{Training} & \textbf{Testing} & \textbf{Inference} \\ 
        \midrule
        SafetyNets~\cite{ghodsi2017safetynets} & 2017.6 & DNN & $\circ$ & $\circ$  & $\bullet$  \\
        Drynx~\cite{froelicher2020drynx} & 2020.3 & Regression models & $\circ$ & $\circ$ & $\bullet$  \\
        ZKDT~\cite{zhang2020zero} & 2020.10 & DT & $\circ$ & $\bullet$ & $\circ$  \\
        vCNN~\cite{lee2024vcnn} & 2020.12 & CNN & $\circ$ & $\circ$ & $\bullet$  \\
        ZEN~\cite{feng2021zen} & 2021.5 & NN &  $\circ$   & $\bullet$  & $\circ$ \\
        Mystique~\cite{weng2021mystique} & 2021.8 & NN & $\circ$ & $\circ$ & $\bullet$  \\
        VeriML~\cite{zhao2021veriml} & 2021.10 & Six models & $\bullet$  & $\circ$ & $\circ$\\
        zkCNN~\cite{liu2021zkcnn} & 2021.11 & CNN & $\circ$ & $\bullet$ & $\circ$  \\
        Hydra~\cite{zhang2021hydra} & 2021.12 & NN & $\bullet$  & $\circ$  & $\circ$ \\
        Kang~\cite{kang2022scaling} & 2022.10 & MobileNet v2 & $\circ$ & $\circ$ & $\bullet$  \\
        Singh~\cite{singh2022zero} & 2022.10 & DT & $\circ$ & $\circ$ & $\bullet$ \\
        zkMLaaS~\cite{huang2022zkmlaas} & 2022.12 & ML & $\bullet$  & $\circ$ & $\circ$  \\
        ezDPS~\cite{wang2022ezdps} & 2022.12 & ML pipeline & $\circ$ & $\circ$ & $\bullet$ \\
        pvCNN~\cite{weng2023pvcnn} & 2023.3 & CNN & $\circ$  & $\bullet$  & $\circ$  \\
        Fan~\cite{fan2023validating} & 2023.5 & CNN & $\circ$    & $\circ$  & $\bullet$  \\
        ZKDL~\cite{sun2023zkdl} & 2023.7 & DNN & $\bullet$ & $\circ$   &  $\circ$    \\
        ZKPoT~\cite{garg2023experimenting} & 2023.9 & LogR & $\bullet$  & $\circ$   &  $\circ$  \\
        MSCProof~\cite{balbas2023modular} & 2023.10 & CNN / image processing & $\circ$ & $\circ$ & $\bullet$ \\
        South~\cite{ganescu2024trust} & 2024.2 & ML & $\circ$    & $\circ$  & $\bullet$  \\
        Lookup arguments~\cite{campanelli2024lookup} & 2024.4 & DT & $\circ$   &  $\bullet$  &  $\circ$   \\
        ZKML~\cite{chen2024zkml} & 2024.4 & ML & $\circ$    & $\circ$   & $\bullet$ \\
        zkLLM~\cite{sun2024zkllm} & 2024.4 & LLM & $\circ$    & $\circ$ & $\bullet$  \\
        Lu~\cite{lu2024efficient} & 2024.5 & NN & $\circ$    & $\circ$   & $\bullet$  \\
        SpaGKR~\cite{li2024sparsity} & 2024.6 & ML & $\circ$     & $\circ$ & $\bullet$   \\
        Kaizen~\cite{abbaszadeh2024zero} & 2024.7 & DNN & $\bullet$  & $\circ$   & $\circ$   \\
        Hao~\cite{hao2024scalable} & 2024.8 & ML & $\circ$   &  $\circ$ & $\bullet$ \\
        Artemis~\cite{lycklama2024artemis}    & 2024.9 & ML & $\circ$  & $\bullet$ &  $\circ$  \\
        Zhan~\cite{zhan2024validating}   & 2024.11 & CNN & $\circ$  & $\circ$  &  $\bullet$  \\
        zkGPT~\cite{qu2025zkgpt}   & 2025.06 & LLM & $\circ$  & $\circ$  &  $\bullet$  \\ 
        VeriLoRA~\cite{liao2025zklora}   & 2025.08 & LLM & $\bullet$  & $\circ$  &  $\circ$  \\
        \bottomrule
    \end{tabular}}
    \label{tab3}
\end{table*}

While zkPoT~\cite{garg2023experimenting} demonstrates that proof-of-training can be practical for relatively simple models such as logistic regression, it does not directly address the scale and structure of deep neural network training. To bridge this gap, Abbaszadeh \emph{et al.} proposed a dedicated proof-of-training framework for DNNs, referred to here as Kaizen~\cite{abbaszadeh2024zero}. Their design targets small-batch gradient descent and centers on two ideas: an optimized GKR-style proof system for verifying each gradient-descent iteration, and a recursive composition framework for aggregating proofs across many iterations into a single concise proof. At the protocol level, Kaizen tailors a sumcheck-based construction to the algebraic structure of gradient updates, thereby reducing the proving overhead of each step. At the system level, it combines these per-iteration proofs with polynomial commitments to obtain compact proofs and commitments over long training trajectories. This line of work marks an important extension of zkPoT-style thinking from shallow models to modern deep learning workloads.

VeriLoRA extends verifiable training to large language models, with particular emphasis on LoRA-based fine-tuning~\cite{liao2025zklora}. Unlike earlier systems that focus mainly on DNNs or CNNs, VeriLoRA offers an end-to-end framework that covers forward propagation, backward propagation, and LoRA weight updates within transformer-based models. For arithmetic-intensive matrix operations, it employs sumcheck protocols over multilinear extensions; for non-arithmetic operations such as SwiGLU, Softmax, element-wise products, and transposition, it adopts specialized lookup-based subprotocols. The system is evaluated on OPT and LLaMA-family models ranging from 3B to 13B parameters. Reported results indicate proving times below 600 seconds per sample, verification times below 4 seconds, and peak GPU memory below 80GB through layer-wise sequential execution. More broadly, VeriLoRA shows that modern parameter-efficient LLM fine-tuning can be brought into the scope of verifiable training without abandoning practical efficiency.

\subsubsection{\bf ZKP-based Verifiable Testing}\label{sss:3b}

In 2020, \cite{zhang2020zero} proposed a ZKML protocol for verifiable inference and accuracy testing of decision trees, named zkDT. zkDT enables the owner of a decision tree model to convince others of the model's inference results on data samples or its accuracy on public datasets, all without revealing any information about the model itself. The protocol leverages the Aurora protocol~\cite{ben2019aurora} as the ZKP backend due to Aurora's fast proving time, which is a desirable feature for large decision trees. zkDT significantly improves the efficiency of proving time by transforming the computation of decision tree inference into a circuit of size $O(d + h)$, where $d$ is the length of the inference path on the tree, and $h$ is the number of data features. Notably, many testing data samples share common nodes on the decision tree. To exploit this, zkDT optimizes the proof and verification process for decision tree accuracy testing by validating all the nodes of the inference paths across all testing data in a single step, rather than validating the inference path of each sample individually. The implementation of zkDT demonstrates its practicality. For a decision tree with 23 levels and 1029 nodes, and a test dataset consisting of 5000 data samples with 54 features each, zkDT takes 250 seconds to generate a proof of size 287 KB for accuracy testing, and 15.6 seconds for verification.

\cite{campanelli2024lookup} later revisited decision-tree verification through the lens of lookup arguments and showed how the original zkDT construction can be substantially improved. In general, lookup arguments allow the prover to show that committed values originate from a larger committed table, which is especially attractive for non-arithmetic operations such as range checks, XOR, and AND. Building on this idea, the authors extend vector lookups to matrix lookups, introduce privacy-preserving variants for both the queried substructure and the underlying table, and design more efficient lookup arguments such as cq+, zkcq+, and cq++~\cite{eagen2022cq}. Their application to decision trees is particularly elegant: the tree is encoded as a committed matrix, while the prover commits only to the row corresponding to the reached leaf and then proves that this row belongs to the committed matrix. The prover can subsequently show that the input satisfies the predicates encoded in that row. For batched statements, multiple reached leaves can be handled through a committed matrix of rows. Because the resulting matrix-lookup protocol removes the dependence on tree size from the prover's asymptotic complexity, it yields a markedly cleaner and more scalable approach to verifiable testing of decision trees.

pvCNN~\cite{weng2023pvcnn} targets verifiable testing of CNNs through a hybrid design that combines homomorphic encryption, collaborative inference, and zk-SNARKs. Its core technical contribution is the use of Quadratic Matrix Programs (QMPs) to encode convolution more efficiently and to support aggregation of proofs across multiple test instances. The framework further splits the model into private and public components, allowing encrypted partial execution while still producing verifiable test results. Relative to earlier QAP-based approaches, pvCNN substantially reduces setup and proving cost for CNN testing.

ZEN~\cite{feng2021zen} can be viewed as one of the first optimizing compilers explicitly designed for verifiable neural networks. It provides two schemes, ZENacc for proving model accuracy and ZENinfer for proving individual inference results, and it improves efficiency mainly through proof-friendly quantization and stranded encoding. These optimizations reduce the R1CS cost of neural operators while preserving model accuracy, making ZEN an important step from hand-crafted circuits toward compiler-assisted ZKML development.

\cite{liu2021zkcnn} proposed a scheme for Convolutional Neural Networks (CNNs), called zkCNN, which enables the verification of a model's inference for a given input data without revealing the model's parameters. This scheme can also be generalized to prove the model's accuracy on a public dataset while ensuring that the model's parameters remain private. zkCNN incorporates a new sumcheck protocol for two-dimensional convolutions, achieving a prover time of $O(n^2)$ for two input matrices of sizes $n \times n$ and $w \times w$. This is even faster than computing the result directly, making it asymptotically optimal. The proof size is $O(\log n)$. A key component of this protocol is an efficient sumcheck for the Fast Fourier Transform (FFT), which requires only $O(N)$ time to generate the proof for a vector of size $N$, offering a better asymptotic complexity than the conventional $O(N \log N)$ needed for FFT computation. Building upon this, the authors further propose a protocol that achieves a sublinear verifier time of $O(\log^2 n)$ with a proof size of $O(\log^2 n)$. Additionally, they design an interactive proof using the Generalized Knowledge Representation protocol for CNN predictions, including verification of the convolutional layers, ReLU activation functions, and max pooling operations. They introduce generalized addition and multiplication gates, which allow operations with fan-in greater than two, enabling inner products to be implemented with a single sumcheck. By extending these gates to accept inputs from any layer, they avoid additional prover time. For CNN convolutional layers, they optimize the sumcheck protocol by reducing the prover time for the inverse FFT (IFFT) by a factor corresponding to the number of input channels. They also design an efficient circuit gadget that combines the computation of the ReLU activation and max pooling functions, requiring only a single bit-decomposition per number. The proposed scheme supports large CNNs like VGG16, which has 15 million parameters and 16 layers, significantly improving the proof generation time to 88.3 seconds---1264 times faster than existing schemes. The proof size is 341 kilobytes, and the verifier time is only 59.3 milliseconds. Furthermore, the scheme can scale to prove the accuracy of the same CNN on 20 images.

A separate bottleneck is the cost of checking commitments to models and data inside commit-and-prove pipelines. Apollo and Artemis~\cite{lycklama2024artemis} address this issue by redesigning CP-SNARK linking so that commitment verification places much less burden on the underlying SNARK. Apollo targets KZG-based settings and relies on white-box access to the proof system, whereas Artemis works with generic homomorphic polynomial commitments using only black-box access. This distinction is especially important for ZKML, where commitment checking can dominate the cost of testing or inference proofs on larger models and datasets. In practice, Artemis is particularly attractive because it remains compatible with modern IPA- and Halo2-style settings that avoid trusted setup while still approaching Apollo-level performance.

\subsubsection{\bf ZKP-based Verifiable Inference}\label{sss:3c}

The earliest scheme for verifiable inference based on ZKP is SafetyNets~\cite{ghodsi2017safetynets}, proposed in 2017. This approach enables verifiable inference of deep neural networks on untrusted cloud servers. Noting that the hierarchical structure of the GKR protocol aligns almost perfectly with the architecture of multi-layer neural networks, SafetyNets combines the GKR protocol with an interactive proof protocol for matrix multiplication~\cite{cormode2011verifying}, achieving end-to-end verifiability while significantly reducing bandwidth costs. To address the challenge of nonlinear computations in neural networks, which cannot be easily represented in arithmetic circuits and handled by the proof protocol, SafetyNets restricts its support to specific quadratic activation functions and sum pooling operations. However, this limitation reduces the generality of the technique. Although SafetyNets does not provide zero-knowledge properties, privacy is not a concern in this context, as both the input and the model are supplied by the verifier.

In 2020, \cite{froelicher2020drynx} proposed a decentralized ZKML system named Drynx that combines interactive protocols, homomorphic encryption, zero-knowledge proofs, and differential privacy technologies, enabling privacy-preserving statistical queries (inferences) and training and evaluation of logistic regression models on distributed datasets. The Drynx system consists of four main components: the querier (Q), data providers (DPs), computation nodes (CNs), and verification nodes (VNs). Within the system's interaction protocol, Q initiates a query request, after which the DPs encrypt their raw data using homomorphic encryption techniques and transmit the encrypted responses to the CNs. The CNs aggregate and process the received encrypted data collectively, generating both the query results and corresponding zero-knowledge proofs to ensure the correctness of the computation process. The VNs are responsible for verifying the correctness of the computations. Through this interaction protocol, Drynx effectively guarantees the transparency and verifiability of query execution. Even under a strong adversarial model with malicious entities, the system ensures the correctness of the computation results. Drynx employs Camenisch-Stadler ZKP~\cite{camenisch1997proof} to verify the computational integrity of CNs, and Camenisch-Chaabouni ZKP~\cite{chaabouni2007efficient} to validate the range of input data, ensuring that the data provided by DPs falls within legitimate bounds. CNs generate zero-knowledge proofs upon completing calculations, while VNs collectively verify these proofs to ensure the accuracy of computational results, with verification results stored on the blockchain to achieve computational process auditability. Additionally, the Drynx system implements differential privacy protection based on the Collective Differential Privacy (CDP) protocol introduced in Unlynx~\cite{froelicher2017unlynx}, effectively protecting individual privacy by adding noise to query results.

vCNN is a verifiable inference framework specifically designed to accelerate the proof process for convolutional neural networks (CNNs)~\cite{lee2024vcnn}. In CNNs, the output of a convolutional operation is represented as a sum of products: $y_i = \sum\nolimits_{j = 1}^{l - 0} {a_j \cdot x_{l - 1 - j}}$, where \(y_i\) is the \(i\)-th convolutional output, \(\left( x_i, \cdots, x_{i+l-1} \right)\) is the \(i\)-th input data vector of length \(l\), and \(\left( a_0, \cdots, a_{n-1} \right)\) is the convolutional kernel vector of length \(n\). The number of multiplications involved in a convolutional operation is \(O(ln)\), which is often computationally expensive. Consequently, traditional zk-SNARKs incur significant proving time when verifying these multiplications. To address this challenge, vCNN introduces a novel approach to reduce the cost of proving CNN convolutional operations by transforming the sum of products into a product of sums: $\left( \sum\nolimits_{i = 1}^{n + l - 2} y_i \right) = \left( \sum\nolimits_{i = 0}^{n - 1} x_i \right) \left( \sum\nolimits_{i = 0}^{l - 1} a_i \right)$. However, this transformation introduces a potential issue: multiple different values of \(y_i^\prime \neq y_i\) can satisfy the above equality if \(\sum\nolimits_{i = 1}^{n + l - 2} y_i^\prime = \sum\nolimits_{i = 1}^{n + l - 2} y_i\). To prevent such ambiguities, an indeterminate variable \(Z\) is introduced, ensuring that the following equality holds for all \(Z\): $\left( \sum\nolimits_{i = 1}^{n + l - 2} y_i Z^i \right) = \left( \sum\nolimits_{i = 0}^{n - 1} x_i Z^i \right) \left( \sum\nolimits_{i = 0}^{l - 1} a_i Z^i \right)$. This equality corresponds to a polynomial multiplication: $y(Z) = x(Z) \cdot a(Z)$, where \(y(Z)\), \(x(Z)\), and \(a(Z)\) are polynomials. The verifiability of this transformation can be encoded using quadratic polynomial programs (QPPs)~\cite{kosba2014trueset}, reducing the proving cost from \(O(ln)\) to \(O(n + l)\) when \(n\) and \(l\) take practical values. This reduction in complexity enables vCNN to significantly improve the proving time for CNN inference compared to traditional zk-SNARKs based on quadratic arithmetic programs (QAPs)~\cite{burkard1984quadratic}. Experimental results demonstrate the effectiveness of vCNN. On the simple MNIST dataset, vCNN improves proof performance by a factor of 20. For the more complex VGG16 model~\cite{simonyan2014very}, it achieves an improvement of 18,000 times.

As researchers moved toward larger neural networks, the lack of efficient proof primitives for complex inference workloads became a major bottleneck. Mystique~\cite{weng2021mystique} addresses this issue by building an sVOLE-based~\cite{yang2021quicksilver} framework that provides reusable proof components for neural-network inference. Its main contributions are threefold. First, Mystique offers a collection of practical ZKML building blocks that abstract away much of the low-level cryptographic complexity. Second, it introduces efficient conversion protocols between arithmetic and Boolean values, between public commitments and private authenticated values, and between fixed-point and floating-point representations, all of which are central to bridging ML computations with proof-system arithmetic. Third, it presents an optimized matrix-multiplication proof whose number of private multiplications is sublinear in the matrix size. Taken together, these ideas make Mystique an important systems contribution, especially for later work on more expressive and scalable inference verification.

MSCProof~\cite{balbas2023modular} pushes this modular direction further with a framework for sequential verifiable computation based on Verifiable Evaluation schemes on Fingerprinted Data (VEs), an abstraction that captures diverse sumcheck-style proofs including GKR. It also designs a VE specialized for batched multi-channel convolution and uses it to build modular proofs for CNN and image-processing pipelines. This work is particularly relevant to ZKML because it narrows the gap between highly tailored ML proof systems and more composable proof-engineering workflows.

\cite{singh2022zero} proposed a zero-knowledge verifiable scheme for a distributed AI pipeline, which includes a privacy-preserving verification protocol for inference using decision trees. The distributed AI pipeline assigns different stages of the process---such as data collection, model training, and model inference---to independent participants, including data owners, model owners, and model users. However, previous approaches faced significant inefficiencies in handling memory within the circuit. To overcome this limitation, the authors introduced an improved protocol for the privacy-preserving verifiable inference using decision trees. By eliminating the need for expensive one-time hash operations on the tree structure, the size of the verification circuit was reduced by up to tenfold. Additionally, the protocol leverages read-only memory access to further optimize performance, significantly reducing the number of multiplication gates required per prediction. Experimental results demonstrate that, compared to zkDT---a similar scheme designed for the verification of decision tree inference---the proposed protocol achieves substantial efficiency improvements, with its circuit size being only $25\%$ to $40\%$ of zkDT.

In December 2022, \cite{wang2022ezdps} proposed ezDPS, an efficient machine learning inference pipeline (MLIP) designed to process data across multiple stages at the ML service provider's end. The pipeline enables the service provider, who supplies the ML model, to compute the final inference result for the client, who provides the input data, while safeguarding the private model parameters at each stage. After committing to its private model, the ezDPS MLIP operates through three key processing stages: (i) utilizing discrete wavelet transform (DWT) for initial data transformation; (ii) applying principal component analysis (PCA) to reduce dimensionality and extract essential features; (iii) performing classification using support vector machines (SVM). The framework employs Hyrax~\cite{wahby2018doubly} as the underlying polynomial commitment scheme and Spartan~\cite{setty2020spartan} as the backend zero-knowledge proof (ZKP) protocol to verify computations across these stages. Furthermore, ezDPS introduces the concept of zero-knowledge proof of accuracy (zkPoA), which allows the ML service provider to prove the accuracy of the committed model on public datasets without revealing the model parameters. However, a notable limitation of this approach is its inability to preserve client data privacy. Specifically, clients must send their input data in plaintext to the ezDPS platform for computation, potentially exposing sensitive information.

\cite{halo2book} leveraged the Halo2 ZKP scheme to construct a zk-SNARK arithmetic circuit capable of supporting the MobileNet v2 model~\cite{sandler2018mobilenetv2} in the paper \cite{kang2022scaling}. In contrast to prior works that primarily targeted small-scale datasets (\emph{e.g.}, MNIST or CIFAR-10) with simpler models, this study demonstrates that zk-SNARKs can effectively handle real-world, large-scale neural network models. Specifically, it supports MobileNet v2, a model trained and evaluated on the large-scale ImageNet dataset---a widely used computer vision benchmark comprising over 1 million high-resolution images across 1,000 categories. This work primarily addresses the computational overhead of verifying division operations in the circuit, which is a significant bottleneck in zk-SNARK-based deep learning inference. To tackle this, the authors proposed two key optimizations based on the Plonkish arithmetization framework~\cite{gabizon2019plonk} in the Halo2 scheme. (1) For linear operations, including convolutional layers, residual connection layers, and fully connected layers, the authors designed two custom gates to efficiently encode division operations. The first gate performs an addition of multiple inputs, while the second gate computes the dot product of inputs and weights, incorporating the division by a fixed scaling factor. These gates avoid the need for complex floating-point arithmetic by leveraging fixed-point approximations, significantly reducing the number of constraints required for division. (2) For nonlinear operations such as ReLU and softmax, the authors utilized lookup arguments to efficiently represent division. Specifically, the division operation is precomputed for a range of possible input values, and the results are stored in a lookup table. During proof generation, the circuit enforces that the computed division results match the precomputed values in the table, thereby reducing the computational cost associated with division. Additionally, the authors ensured the privacy of the model's inputs and weights by incorporating hash commitments. This allows the model provider to prove the correctness of inference without revealing sensitive model parameters or input data. The proposed zk-SNARK scheme achieves high accuracy on ImageNet-scale models while maintaining relatively low verification time. For example, the proof verification time for MobileNet v2 with 79.2\% accuracy on ImageNet is approximately 10.27 seconds on commodity hardware. Furthermore, the proof size is significantly smaller compared to prior methods. Specifically, the proof size is reduced to 5,952 bytes, which is orders of magnitude smaller than methods based on secure multi-party computation (MPC), which typically require tens to hundreds of gigabytes for proof representation.

\cite{fan2023validating} proposed a zero-knowledge verification scheme to ensure the integrity of CNN inference, with a focus on balancing the confidentiality of certain CNN model parameters and verifying the correctness of the inference process in machine learning as a service (MLaaS) scenarios. In this scheme, the authors developed a computational logic extraction algorithm capable of accurately translating the computational logic of convolutional layers, fully connected layers, pooling layers, and activation layers into corresponding simple arithmetic expressions. These expressions are subsequently used to construct proof circuits for each layer of the model within the zk-SNARK framework. Based on this design, the prover performs the computations involved in the CNN model's inference process and generates zero-knowledge proofs using the constructed proof circuits. The verifier then validates the integrity of the inference process by checking the proofs for each layer of the model. Although this scheme increases the number of multiplications and additions, resulting in a more complex R1CS for circuit construction, experimental results show that both the computational and storage overheads remain within acceptable limits. This demonstrates the practicality of the proposed approach for real-world applications.

\cite{ganescu2024trust} studied Transformer-scale proving through snarkGPT, an implementation built on Halo2 and nanoGPT~\cite{karpathy2023nanogpt}. Rather than treating LLM proving as a straightforward extension of CNN-style pipelines, the paper highlights how proving cost grows rapidly with layer width, depth, and circuit matrix size, with memory quickly becoming a dominant constraint. Its main value in the survey is therefore diagnostic: it makes clear that Transformer verification requires different engineering assumptions from earlier small-model ZKML systems.

\cite{chen2024zkml} moves closer to large-model practical deployment by introducing a Halo2-based compiler for TensorFlow models together with efficient low-level gadgets and a circuit-layout optimizer. The system supports a broader range of architectures, including vision models and smaller Transformer variants, while improving proving time, verification time, and proof size relative to earlier frameworks. More importantly, it shows that compiler quality and layout optimization are now as important as proof-system choice for practical ZKML inference.

zkLLM~\cite{sun2024zkllm} extends verifiable inference to large language models through two main components: tlookup for lookup-based handling of nonlinear tensor operations, and zkAttn for more efficient verification of attention and Softmax-related computation. By exploiting lookup compression and GPU parallelism, it demonstrates that LLM inference proofs can be generated for models up to 13B parameters within practical memory and time budgets. This makes zkLLM a milestone in pushing ZKML beyond CNN-scale inference.

Lu \emph{et al.}~\cite{lu2024efficient} propose a complementary inference framework built around VOLE, range proofs, and lookup proofs. Their design focuses especially on reducing the cost of nonlinear layers and keeping the system modular enough to support both CNNs and Transformer models. Compared with earlier bit-decomposition-heavy approaches, this line of work shows that alternative proving backends can yield large efficiency gains while preserving support for higher-precision quantized networks.



\cite{hao2024scalable} conducted an in-depth study on improving the efficiency of ZKP for non-linear functions in ML models. They proposed the use of the table lookup technique and digital decomposition to address the computational cost issues in verifying non-linear functions. Their specific contributions are as follows. (1) {\bf Table Lookup Technique}: This is one of the core innovations of the paper, aiming to efficiently verify the correctness of non-linear functions through the use of lookup tables. Specifically, complex non-linear functions (such as exponentials, divisions, etc.) are transformed into lookup table problems by storing valid input-output mappings in a lookup table. The prover needs only to prove that the input-output pair exists in the lookup table to complete the function verification. This method avoids the traditional verification approach based on Boolean circuits, eliminating the need to convert arithmetic values into Boolean values, thereby significantly reducing computational complexity. (2) {\bf Digital Decomposition}: However, directly storing lookup tables of all possible input-output pairs leads to excessively large sizes, especially when the input bit-length is large, causing the size of the lookup table to grow exponentially. To solve the problem of lookup table size explosion, the paper proposes digital decomposition. Large bit-length inputs cause storage requirements and computational costs of lookup tables to increase sharply; therefore, the paper decomposes large bit-length inputs into several smaller numbers (\emph{e.g.}, 5 to 12 bits), effectively reducing the size of the lookup table. For example, an input~$x$ can be decomposed as $x = x_{k-1} \parallel \ldots \parallel x_0$, where each decomposed number~$x_i$ has a smaller bit length. This approach allows the lookup table to store mappings of small numbers for verification. Through digital decomposition, the paper supports the verification of large bit-length inputs with less storage overhead and computational cost while maintaining the correctness and completeness of the verification results. Through these two key steps, the authors successfully implemented zero-knowledge proofs for common non-linear functions such as ReLU, Sigmoid, GELU, Softmax, Maxpooling, and normalization. Compared to existing solutions (\emph{e.g.}, Mystique~~\cite{weng2021mystique}), this research achieved significant performance improvements, with runtime reduced by $50\times$ to $179\times$, and communication costs decreased by $1.2\times$ to $4.8\times$. In applicability verification, the framework demonstrated outstanding efficiency advantages in mainstream machine learning tasks (such as inference verification of CNNs and LLMs), especially showing significant performance improvements when handling common functions like ReLU and Softmax. For example, on the ReLU function, runtime was reduced by approximately $100\times$; on the Softmax function, runtime was reduced by approximately $179\times$; and on the GELU function, runtime was reduced by approximately $77\times$ to $86\times$.

\cite{li2024sparsity} introduced a ZKP-based verifiable inference framework for deep learning models, aiming to enhance the efficiency and practicality of ZKML while addressing privacy and security concerns in current MLaaS systems. The framework achieves significant improvements in proof efficiency and storage overhead by incorporating sparsity-aware protocols and ternary networks. Its implementation involves the following key steps: (1) Optimized Protocol for Sparse Linear Layers (SpaGKR-LS): The paper proposes an optimized protocol, SpaGKR-LS, specifically designed for the linear layers of neural networks. This protocol achieves proof time that scales linearly with the number of non-zero parameters, significantly reducing computational complexity through techniques such as mode pruning and quantization. By leveraging the sparsity of linear layers, the protocol enables more efficient proofs for sparse networks. (2) Quantization with Ternary Networks: The framework quantizes the model using ternary networks, where parameters are restricted to \{-1, 0, 1\}. This design eliminates multiplication operations and, by combining sparsity and low-bit-width characteristics, further enhances proof efficiency. The ternary network's inherent sparsity and simplified arithmetic operations reduce computational overhead, making the proof process more efficient. (3) Modular Framework Design: A modular framework is proposed, compatible with existing ZKML methods based on GKR (Sumcheck Protocol). This modular design allows seamless integration with sparsity-aware protocols, such as the Lasso protocol~\cite{setty2024unlocking} or the SpaGKR protocol proposed in this work. The flexibility of the modular approach enables efficient verification and adaptability to different use cases. Through these three steps, the proposed framework not only overcomes the efficiency bottlenecks of traditional ZKML methods in handling sparsity and quantized models but also significantly enhances the flexibility and scalability of the framework through its modular design. Experimental results demonstrate the effectiveness of the framework. For sparse linear layers, SpaGKR-LS achieves a 45$\times$ speedup in proof time compared to traditional ZKML methods that ignore sparsity. For ternary networks, proof time is further reduced by approximately 5$\times$. Additionally, the framework achieves substantial improvements in verification time and storage efficiency, showcasing its practicality and effectiveness for large-scale deep learning models.

The work of~\cite{zhan2024validating} proposes a ZKP-based verifiable inference scheme for deep learning models, aiming to address data leakage and service fraud issues prevalent in current MLaaS platforms. By ensuring the integrity of ML inference processes and results, as well as the privacy and security of ML model parameters, the scheme enhances trust in MLaaS. The proposed scheme integrates non-interactive ZKP with blockchain technology to ensure model integrity verification while protecting sensitive information about model parameters. Its implementation can be divided into three key steps: (1) R1CS Circuit Design for Neural Network Modules: Each module of the neural network---such as convolution, normalization, activation, and pooling layers---is designed as an R1CS circuit to precisely describe and verify its computational logic. To optimize the circuit complexity, several innovations are introduced in the module designs: In the \emph{convolutional layers}, \emph{depthwise separable convolutions} are employed instead of traditional convolutions, significantly reducing computational complexity; for the \emph{activation functions}, minimal polynomial methods are used to approximate complex activation functions (\emph{e.g.}, Swish), avoiding the complexity associated with floating-point operations; in the \emph{pooling layers}, \emph{adaptive average pooling} circuits are designed to ensure the circuits can flexibly adapt to different input and output sizes. These optimizations enable the computational logic of each module to be transformed into simple polynomial constraints, significantly reducing the generation time and storage requirements of zk-SNARK proof files. (2) Serial Hash Circuit Connection (SHCC) Algorithm: To address the high complexity introduced by interactions and integrity verification between modules in the modular design of large neural networks, the paper proposes the SHCC algorithm. The large-scale deep learning model is divided into multiple modules, with hash connections ensuring integrity between modules. Specifically, the SHCC algorithm adopts a layer-by-layer hashing method, where each module's output hash value is generated from the current module's computation result and the previous module's hash value, forming a serial hash chain. If the output of any module is tampered with, subsequent module verification will fail due to hash mismatches, thereby ensuring the integrity and security of the entire network. Through these steps, the proposed framework not only ensures the integrity verification of deep learning models but also enhances computational and storage efficiency while guaranteeing privacy protection. Experimental results demonstrate significant optimizations in verification time and storage overhead. Specifically, the verification time is reduced by $54.6\%$. In terms of storage overhead, the transformed R1CS circuits are more concise, reducing the storage requirements of proof files (such as R1CS files, ZKEY files, and WASM files) by $58.1\%$. Notably, the proof files generated using the Groth16 protocol have a fixed size of $4$KB. In the Ethereum environment, verification of proof files is achieved through smart contracts. The verification process is divided into three modules: \emph{input module}, \emph{backbone module}, and \emph{output module}, with proof files generated off-chain separately and verification executed through on-chain smart contracts. Experimental results confirm the feasibility of the scheme through the deployment and execution of smart contracts. Gas consumption tests demonstrate that the verification process is efficient and cost-controllable; for example, the execution cost of the backbone module is 397,434 Gas, and the verification cost is 222,424 Gas.

zkGPT~\cite{qu2025zkgpt} is a recent non-interactive framework for the verifiable inference of large Transformer models such as GPT-2. Building on GKR-style proving, it improves prior systems along three main dimensions: a more efficient matrix-multiplication layer based on grouped multilinear evaluation and precomputation; a ``result-as-witness'' paradigm for nonlinear and normalization operators such as Softmax, LayerNorm, and GeLU; and system-level optimizations, including constraint fusion and circuit squeezing, that reduce range-check overhead and increase parallelism. The resulting system produces succinct Fiat--Shamir proofs of about 101~KB and reduces end-to-end GPT-2 proving time to under 25 seconds. Although it currently focuses on inference under quantized arithmetic and does not yet address verifiable training or long multi-token generation, zkGPT shows how Transformer-scale inference can be pushed much closer to practical ZKML deployment.

Additionally, the research efforts in zkDT~\cite{zhang2020zero}, ZEN~\cite{feng2021zen}, zkCNN~\cite{liu2021zkcnn}, and Lookup arguments~\cite{campanelli2024lookup} also contribute to and support ZKP-based verifiable inference. These works have been introduced earlier in this section and will not be revisited in this part discussing verifiable inference.

\subsection{Implementation Improvements in ZKML}\label{ss:3c}

Despite the rapid progress reviewed above, practical ZKML systems remain difficult to engineer because mainstream ML computation is not naturally aligned with proof systems over finite fields. Two bottlenecks recur throughout the literature:

\begin{itemize}
    \item \emph{Generality limitations}: Most ML models rely on floating-point arithmetic, nonlinear activations, and comparison-style logic, whereas many practical ZKP systems operate over finite-field arithmetic circuits. As a result, common ML operators cannot be translated directly into proof-friendly representations. ReLU, for example, involves comparison and conditional behavior, so it must be encoded through auxiliary binary variables, polynomial approximation, or other operator-specific gadgets. These adaptations rarely transfer cleanly across models or proof systems. In practice, each new activation, normalization rule, or quantization strategy often requires custom arithmetization, precision analysis, and optimization.


    \item \emph{Efficiency barriers}: Even after successful arithmetization, modern ML models induce enormous circuits. Large neural networks lead to high proving time, large commitment and key material, and substantial memory pressure. A naive zk-SNARK treatment of VGG16-scale inference, for instance, would yield an impractically large circuit and correspondingly prohibitive prover cost. Similar bottlenecks arise in witness generation, commitment verification, and proof composition. Reducing end-to-end proof cost without weakening soundness therefore remains a central implementation challenge.
    
\end{itemize}

\begin{figure*}[t]
    \centering
    \includegraphics[width=\linewidth]{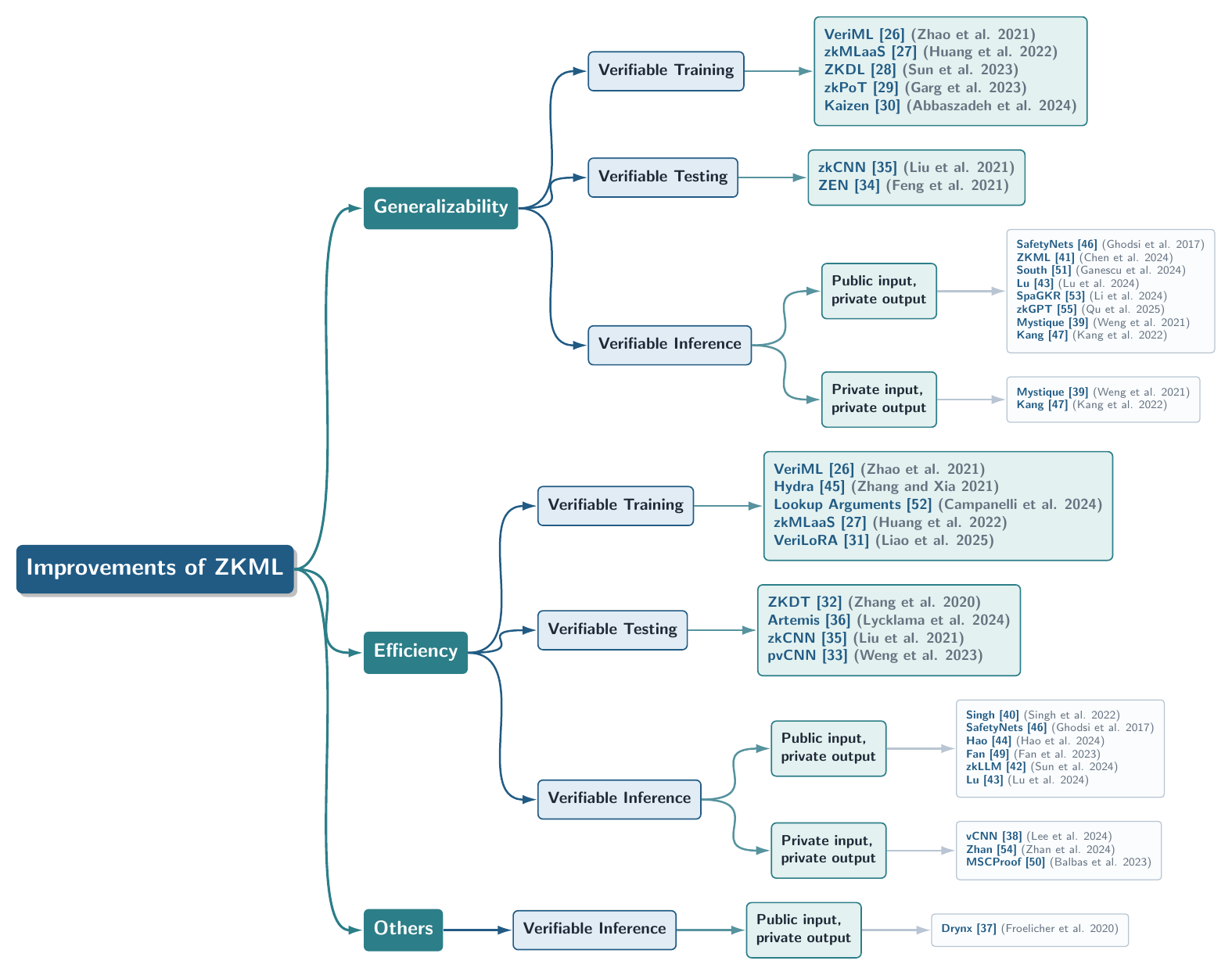}
    \caption{Taxonomy of representative ZKML studies organized by the main type of improvement they target. The figure first groups prior work into improvements in generalizability, improvements in efficiency, and other system-level directions. Within each branch, representative studies are further organized by verification objective, including verifiable training, verifiable testing, and verifiable inference; inference-oriented work is additionally separated by privacy setting, such as public-input/private-output and private-input/private-output scenarios. This taxonomy highlights how existing ZKML systems differ not only in optimization goals, but also in the ML task and privacy model they support.}
    \label{tab}
\end{figure*}

Existing work addresses these issues through two main directions---improving generality and improving efficiency---together with several related system-level optimizations, as summarized in Fig.~\ref{tab}.

\subsubsection{Improving Generality}\label{sss:3a}
A central issue in ZKML is how to bridge real-valued ML computation and finite-field proof arithmetic. The most common first step is quantization: real-valued inputs, activations, and parameters are mapped to integers so that they can be processed inside arithmetic circuits. The design tension is clear. Too few bits increase overflow risk and quantization error, whereas too many bits inflate circuit size and proving cost.

Early systems mainly rely on task-specific quantization strategies. SafetyNets rescales weights and inputs with global constants so that all values remain within the valid field range~\cite{ghodsi2017safetynets}. VeriML converts fixed-precision decimals into integers by scaling inputs with powers of two~\cite{zhao2021veriml}. zkCNN adopts affine quantization of the form \(a=L(q-Z)\), where \(L\) is the quantization scale and \(Z\) is the zero point~\cite{liu2021zkcnn,Jacob_2018_CVPR}. Likewise, the Halo2-based MobileNet v2 implementation in~\cite{kang2022scaling} uses low-bit quantized parameters to avoid expensive floating-point arithmetic. These designs make ML computation more compatible with finite-field proof systems, but they remain tightly coupled to particular proof frameworks and model families. Thus, they improve practicality without fully resolving the generality problem.

More general solutions aim to provide reusable conversion and arithmetization primitives. Mystique offers efficient conversions between arithmetic and Boolean values, between public commitments and private authenticated values, and between fixed-point and floating-point representations, together with an optimized matrix-multiplication proof whose private multiplication cost is sublinear in the matrix size~\cite{weng2021mystique}. ZEN introduces quantization-oriented optimizations such as sign-bit grouping and remainder-based verification, allowing floating-point PyTorch models to be compiled more efficiently into finite-field circuits~\cite{feng2021zen}. MSCProof~\cite{balbas2023modular} further contributes a modular composition framework based on verifiable evaluation schemes, showing how reusable sumcheck-style components can be composed across sequential ML and image-processing pipelines. The Halo2-based system of~\cite{kang2022scaling} similarly uses custom gates and lookup arguments to support ImageNet-scale MobileNet v2. Collectively, these works move ZKML away from purely ad hoc operator rewriting and toward more reusable proof-friendly ML representations and more composable verification pipelines.

Nonlinear operators remain another major source of generality loss. SafetyNets circumvents the issue by restricting networks to quadratic activations and sum pooling~\cite{ghodsi2017safetynets}. VeriML and zkMLaaS instead rely on polynomial approximation, including Remez-style approximations and least-squares fitting for common activations~\cite{zhao2021veriml,huang2022zkmlaas,chen2020rosetta}. zkCNN computes ReLU more directly through bit decomposition, while Fan \emph{et al.} express ReLU and Softmax in matrix-oriented forms that are easier to verify~\cite{liu2021zkcnn,fan2023validating}. zkDL goes further with zkReLU, a protocol specialized for ReLU-related forward and backward propagation that preserves tensor structure while reducing proof cost~\cite{sun2023zkdl}. More recently, zkGPT avoids full arithmetic emulation of certain nonlinear operators by treating their outputs as advice and verifying them through inverse relations together with range and lookup checks~\cite{qu2025zkgpt}. This progression reflects a broader trend: recent systems increasingly combine approximation, lookup, and advice-based verification in order to balance generality, accuracy, and proof efficiency.

\subsubsection{Improving Efficiency}
Practical ZKML depends not only on expressive arithmetization, but also on keeping prover time, memory footprint, and commitment overhead within manageable ranges. Existing work improves efficiency through several recurring strategies: rewriting ML operators into proof-friendly algebraic forms, exploiting model structure for parallelism or probabilistic checking, and reducing the overhead of commitments and proof composition.

At the operator level, SafetyNets pioneered GKR-based verification for layered neural computations, but its restriction to quadratic activations and sum pooling limited practicality~\cite{ghodsi2017safetynets}. zkCNN advances this line by designing FFT-aware GKR protocols for two-dimensional convolutions, achieving prover complexity \(O(n^2)\) and substantial empirical speedups over vCNN and ZEN on LeNet~\cite{liu2021zkcnn}. Fan \emph{et al.} similarly recast convolution, pooling, ReLU, and Softmax into matrix-style computations and then apply Freivalds' algorithm to reduce setup and proof costs~\cite{fan2023validating}. Related work in~\cite{zhan2024validating} further lowers circuit complexity through depthwise separable convolutions, approximate activations, adaptive average pooling, and a modular head--backbone--tail design linked by hash-based integrity checks.

Other systems reduce cost by changing the proof representation of convolutional networks. vCNN extends QAP-based zk-SNARKs to quadratic polynomial programs (QPPs), making convolution proofs more natural while retaining committed-proof links across layers~\cite{lee2024vcnn}. pvCNN pushes this direction further with quadratic matrix programs (QMPs), which reduce multiplication gates for convolution and enable proof aggregation across repeated CNN layers~\cite{weng2023pvcnn}. For large language models, zkLLM introduces tlookup for parallel verification of non-arithmetic tensor operations, and VeriLoRA transfers this lookup-based philosophy to verifiable LoRA fine-tuning~\cite{sun2024zkllm,liao2025zklora}.

Efficiency challenges also arise in models beyond CNNs. For decision trees, zkDT reduces path-verification cost through carefully designed sibling information, while later work by Singh \emph{et al.} and Campanelli \emph{et al.} further reduces the cost of tree commitment and leaf verification through improved tree encodings and matrix lookup arguments~\cite{zhang2020zero,singh2022zero,campanelli2024lookup}. These advances make proving time less sensitive to tree size and significantly improve verification performance.

At the system level, several works cut overhead by proving less or by composing proofs more effectively. VeriML lowers cost by committing to intermediate states and proving only verifier-selected training rounds~\cite{zhao2021veriml}. Hydra partitions neural-network training circuits into subcircuits that can be proved and verified in parallel, while also using progressive quantization to limit accuracy loss~\cite{zhang2021hydra}. Apollo and Artemis reduce the commitment overhead of commit-and-prove SNARK pipelines, bringing the commitment-check overhead for models such as VGG much closer to the cost of the underlying proof system~\cite{lycklama2024artemis}. Finally, zkMLaaS combines random sampling with im2col conversion and Freivalds-based matrix verification, yielding an approximately \(273\times\) reduction in proof-generation overhead relative to a more direct zk-SNARK approach~\cite{huang2022zkmlaas}.

\section{Commercial Applications of ZKML}\label{s:4}

Beyond academic prototypes, a small but growing commercial ecosystem has emerged around ZKML. Current efforts fall into two broad categories: enabling toolchains that lower the barrier to deploying proof-backed ML, and application-facing platforms that use ZKP or ZKML to support trust-minimized decision-making, on-chain agents, or human-verification services.

EZKL is one of the most visible general-purpose toolchains for zkML inference~\cite{cryptoeprint:2020:352}. In practice, models are exported from PyTorch or TensorFlow into ONNX, after which EZKL compiles them into zk-SNARK circuits. The framework also supports associated input formatting, witness generation, and proof creation. With continued engineering optimizations, it can prove MNIST-scale models within a few seconds while using modest memory by current zkML standards, making it important enabling infrastructure for prototypes and hackathon-style deployments.

Although not itself a ZKML platform, Coda illustrates the commercial maturation of recursive zk-SNARK infrastructure that later on-chain ML deployments can build upon~\cite{bonneau2020coda}. Its constant-size blockchain proof shows how large computation histories can be compressed into succinct public proofs, a capability that is highly relevant to future verifiable ML services deployed in decentralized settings.

DeFiChain publicly highlights the use of ZKML-style ideas for AI-assisted fraud detection and credit-risk assessment~\cite{DeFiChain}. The central motivation is to validate model outputs without exposing user data or proprietary model internals, thereby combining financial trust with privacy preservation.

Modulus Labs has been one of the most visible teams pushing on-chain ZKML demonstrations~\cite{ModulusXYZ}. Its examples include RockyBot, an on-chain trading bot whose actions are tied to predictions from recurrent models, and \emph{Leela vs. the World}, which exposes a verified on-chain chess engine. Modulus Labs has also published a benchmarking study comparing the speed and cost of different proving systems across model sizes~\cite{moduluslabs2023cost}, helping frame the engineering trade-offs of practical ZKML deployment.

Giza focuses on trustless on-chain model deployment~\cite{Giza}. Its stack combines ONNX for model representation, a transpiler from ONNX to Cairo, an ONNX-Cairo runtime for deterministic execution, and smart-contract components for deployment and interaction. This pipeline illustrates how model serving, proof-oriented execution, and blockchain integration can be combined into a unified deployment workflow.

ZKaptcha targets bot resistance in Web3 environments by providing CAPTCHA-style proof services for smart contracts~\cite{zkaptcha}. Its current implementation relies mainly on ZKP rather than full ZKML, but it points to a natural adjacent application area: proof-backed behavioral classification that can distinguish humans from automated agents without exposing sensitive interaction data. Taken together, these examples suggest that early commercial adoption of ZKML is clustering around on-chain inference, verifiable agent logic, and trust-minimized user interaction.

\section{Hybrid Cryptographic Frameworks for Verifiable ML}\label{s:Hyb}

In practice, ZKP is often most effective when combined with complementary privacy-enhancing technologies rather than used in isolation. Homomorphic encryption, differential privacy, federated learning, and secure multiparty computation each address a different part of the trust problem, but none of them alone provides full computational verifiability. Hybrid designs therefore aim to combine privacy, integrity, and scalability by assigning each primitive the role it handles best. This section highlights representative examples of such combinations.

\subsection{Hybridization of Homomorphic Encryption and ZKP}  

pvCNN~\cite{weng2023pvcnn} illustrates how homomorphic encryption and ZKP can play complementary roles in verifiable inference. Homomorphic encryption protects sensitive intermediate computation, while the proof system attests that the returned result is consistent with the prescribed CNN execution. Drynx~\cite{froelicher2020drynx} shows a broader hybrid pattern in distributed analytics by combining homomorphic encryption, differential privacy, and ZKP so that confidentiality, secure aggregation, and verifiability are handled by different layers of the system rather than by a single primitive.

\subsection{Hybridization of Differential Privacy and ZKP}
Verifiable differential privacy~\cite{wei2025verifiable} targets a different trust gap: even if a mechanism claims to satisfy differential privacy, users still need assurance that the stated noise distribution, sensitivity bound, and privacy budget were actually respected. The scheme therefore couples differential-privacy computation with ZKP-based verification of key steps such as noise generation and parameter compliance. Its main technical challenge is that common differential-privacy mechanisms are not naturally proof-friendly, so the work relies on circuit-oriented linearization techniques to express them within arithmetic constraints.

\subsection{Hybridization of Federated Learning and ZKP}

VPFL~\cite{ma2024vpfl} shows how federated learning can be strengthened by adding both privacy-preserving aggregation and verifiability. In this design, local updates are protected during aggregation with homomorphic encryption, while SNARK-style proofs attest that the server carried out the prescribed federated-averaging computation. The result is a hybrid architecture in which confidentiality of client updates and correctness of aggregation are handled separately but coherently. A complementary perspective is provided by Wang \emph{et al.}~\cite{wang2025zkfl}, who organize zero-knowledge federated learning into a structured ZK-FL framework spanning different FL stages and tasks, and further propose Veri-CS-FL, in which clients prove the quality of local-model performance metrics before server-side selection. This line of work is useful because it highlights that ZKP can support not only aggregation integrity, but also upstream decision points such as trustworthy client selection.

\subsection{Hybridization of Secure Multiparty Computation and ZKP}

CrypTen~\cite{knott2021crypten} is best viewed as a representative MPC substrate rather than a ZKP-integrated system. Its relevance here is that it clarifies the role MPC can play in hybrid verifiable ML: MPC can protect multi-party training or inference over private inputs, while ZKP can be layered on top to attest the correctness of selected computations, aggregates, or model-update rules. This separation of roles is appealing because it avoids forcing a single primitive to provide both high-throughput privacy and full public verifiability.

\subsection{Benefits of Hybrid Cryptographic Designs}

The integration of zero-knowledge proofs with complementary cryptographic techniques offers several advantages over standalone approaches:
\begin{itemize}
    \item \textbf{Computational efficiency}: Hybrid approaches delegate bulk confidential computation to primitives such as HE or MPC while reserving ZKP for correctness checks. For example, pvCNN shows that separating confidential computation from proof generation can reduce end-to-end overhead.
    \item \textbf{Security completeness}: Multi-technology fusion covers multiple threat dimensions at once. Drynx, for instance, combines differential privacy, homomorphic encryption, and ZKP to address individual privacy, computation confidentiality, and protocol integrity simultaneously.
    \item \textbf{Scalability}: Carefully chosen combinations can support larger deployments than any single primitive alone. VPFL demonstrates this point by maintaining logarithmic verification overhead as the number of participants grows.
    \item \textbf{Reduced trust assumptions}: Hybrid designs can remove trust assumptions left open by individual technologies. Examples include verifiable differential privacy, which reduces reliance on honest implementation of privacy parameters, and MPC-based ML pipelines whose critical outputs can be attested with ZKP.
\end{itemize}

\section{Future Directions for ZKML}\label{s:5}

Although ZKML has advanced rapidly in recent years, the field is still far from mature. Based on the current research landscape and the remaining technical bottlenecks, we highlight five especially important directions for future work.
 \subsection{Improving Computational Efficiency and Scalability}
Although recent systems have substantially improved performance, efficiency and scalability remain the most immediate barriers to practical deployment. This is especially true for large neural networks, multimodal models, and generative systems, whose circuit sizes, witness-generation costs, and memory footprints can still be prohibitive.

Future work should therefore continue to explore recursive proof composition, more efficient circuit layouts, hardware-aware witness generation, and proof systems tailored to layered ML computation. Progress along these lines will be essential if ZKML is to move from small and medium models toward genuinely large-scale deployment.

\subsection{Improving Generality and Practicality}

Generality and usability remain another major bottleneck. Many current systems are tightly coupled to specific proof backends, quantization rules, operator sets, or model families, and they often require substantial cryptographic expertise to use effectively.

Future progress will likely depend on better compilers, reusable operator libraries, domain-specific languages, and interfaces that fit naturally into mainstream ML toolchains. More standardized development stacks would lower the entry barrier for practitioners. They would also make ZKML easier to evaluate, compare, and deploy across application domains.

\subsection{Support for Diverse and Complex Machine Learning Models}
The scope of supported models also needs to expand. Much of the current ZKML literature still centers on supervised inference or relatively structured training settings, whereas modern ML increasingly involves multimodal pipelines, generative models, retrieval-augmented systems, and self-supervised objectives.

Future research should investigate which components of these systems are most meaningful and tractable to verify. For example, one promising direction is to prove deterministic sub-computations while handling stochastic steps through verifiable randomness or carefully scoped correctness claims.

For self-supervised and unsupervised learning in particular, new theory may be needed to define verification targets that are both cryptographically sound and practically useful.

\subsection{Strengthening Privacy and Security}
Privacy protection is one of the core motivations for ZKML, but existing work still concentrates heavily on inference-time guarantees. Broader end-to-end protection is needed across training, deployment, updating, and model governance.

Future work should therefore study how ZKP can be combined more effectively with MPC, HE, differential privacy, and related techniques to provide stronger privacy guarantees across the full ML lifecycle.

Another open problem is how to reason about robustness and security beyond basic correctness, including adversarial manipulation, model misuse, and malicious updates. Extending ZKML toward these stronger security goals remains challenging but increasingly important.

\subsection{ZKP-Based Verification of Data Authenticity}
Data authenticity is another underexplored direction. Most current ZKML work assumes that the relevant datasets are already legitimate and properly governed, yet in practice the provenance, authorization, and integrity of training or testing data can be just as important as the correctness of the model computation itself.

Future work in this area could include proofs of dataset provenance, authorization, preprocessing integrity, and version consistency, together with scalable mechanisms for large distributed datasets. Bringing such guarantees into ZKML pipelines would significantly strengthen the trustworthiness of privacy-preserving ML services.

\section{Conclusion}\label{s:6}
This survey has reviewed the foundations, representative systems, implementation bottlenecks, commercial efforts, and hybrid cryptographic directions of zero-knowledge machine learning. Across the literature, a clear pattern emerges: ZKML is moving from proof-of-concept demonstrations toward more general, efficient, and practically deployable systems, but substantial challenges remain in scalability, model coverage, privacy across the full ML lifecycle, and data authenticity. We hope this survey provides a useful reference for researchers and practitioners working toward trustworthy, privacy-preserving, and verifiable machine learning services.
  \bibliographystyle{IEEEtranN}
  \bibliography{ref}

\end{document}